\documentclass[twocolumn,preprintnumbers,amsmath,amssymb,showpacs,superscriptaddress,revsymb,prb,footinbib]{revtex4-1}

\usepackage{amssymb,amsmath}
\usepackage{graphicx}
\usepackage{multirow}
\usepackage{subcaption}
\usepackage{hyperref}
\usepackage{appendix}
\usepackage{mathtools}
\usepackage{amsmath}
\usepackage{siunitx}
\graphicspath{{./}}

\begin{document}


\title{First-principles Study of Ce$^{3+}$ doped Lanthanum Silicate Nitride Phosphors: Neutral Excitation, Stokes Shift and Luminescent Center Identification}


\author{Yongchao Jia}
\email[]{yongchao.jia@uclouvain.be}
\affiliation{European Theoretical Spectroscopy Facility, Institute of Condensed Matter and Nanosciences, Universit\'{e} catholique de Louvain, Chemin des \'{e}toiles 8, bte L07.03.01, B-1348 Louvain-la-Neuve, Belgium}
\author{Anna  Miglio}
\affiliation{European Theoretical Spectroscopy Facility, Institute of Condensed Matter and Nanosciences, Universit\'{e} catholique de Louvain, Chemin des \'{e}toiles 8, bte L07.03.01, B-1348 Louvain-la-Neuve, Belgium}
\author{Samuel Ponc\'{e}}
\affiliation{European Theoretical Spectroscopy Facility, Institute of Condensed Matter and Nanosciences, Universit\'{e} catholique de Louvain, Chemin des \'{e}toiles 8, bte L07.03.01, B-1348 Louvain-la-Neuve, Belgium}
\author{Xavier Gonze}
\affiliation{European Theoretical Spectroscopy Facility, Institute of Condensed Matter and Nanosciences, Universit\'{e} catholique de Louvain, Chemin des \'{e}toiles 8, bte L07.03.01, B-1348 Louvain-la-Neuve, Belgium}
\author{Masayoshi Mikami}
\affiliation{MCHC R$\&$D Synergy Center, Inc., 1000,
Kamoshida-cho Aoba-ku, Yokohama, 227-8502, Japan}


\date{\today}

\begin{abstract}
We study from first principles two lanthanum silicate nitride compounds, LaSi$_{3}$N$_{5}$ and
La$_{3}$Si$_{6}$N$_{11}$, pristine as well as doped with Ce$^{3+}$ ion, in view of explaining their different emission color, and characterising the luminescent center. The electronic structures of the two undoped hosts are similar, and do not give a hint to quantitatively describe such difference. The $4f\rightarrow 5d$ neutral excitation of the Ce$^{3+}$ ions is simulated through a constrained density-functional theory method coupled with a ${\Delta}$SCF analysis of total energies, yielding absorption energies. Afterwards, atomic positions in the excited state are relaxed, yielding the emission energies and Stokes shifts. 
Based on these results, the luminescent centers in LaSi$_{3}$N$_{5}$:Ce and La$_{3}$Si$_{6}$N$_{11}$:Ce are identified.
The agreement with the experimental data for the computed quantities is quite reasonable and explains the different color of the emitted light. Also, the Stokes shifts are obtained within 20\% difference relative to  experimental data. 
\end{abstract}


\maketitle

\section{Introduction}
\label{intro}
Phosphors are essential components of white light-emitting-diodes (LEDs). In particular, phosphors activated by the Ce$^{3+}$ ion, presenting a $4f\rightarrow 5d$ spin-allowed transition, have attracted much attention from academia and industry. A large effort has been
devoted to the development of novel Ce$^{3+}$ doped systems in the last
decades.\cite{Hoppe2009,Zeuner2011,Jia2012,Krings2011,Setlur2010,Suehiro2011} 
However, most of these have been found thanks to a semi-empirical
method,\cite{Dorenbos2000a,Dorenbos2000b,Dorenbos2002} that can only provide trends and qualitative predictions. Especially, the prediction on emission property and Stokes shift is limited since the structure geometry of excited state is difficult to measure experimentally. A typical example
beyond the expectation of the semi-empirical method is the emission color of
Ce$^{3+}$ ion in two closely related lanthanum silicon nitrides (LSN),
LaSi$_{3}$N$_{5}$ and
La$_{3}$Si$_{6}$N$_{11}$. In particular, the latter is a blue-convertible yellow phosphor,
with a great potential to replace commercial YAG:Ce, while the former gives 
a blue emission under the UV excitation.

The crystal structure of LaSi$_{3}$N$_{5}$ was studied by Inoue in
1980.\cite{Inoue1980} 
Its structure is built up of
SiN$_{4}$ tetrahedra, which are linked by shared corners.  In this compound, there is only one non-equivalent La$^{3+}$ crystallographic site. The crystal structure of Ln$_{3}$Si$_{6}$N$_{11}$ (Ln: lanthanide element) was reported by Woike and Jeitschko in 1995,\citep{Woike1995} and that of La$_{3}$Si$_{6}$N$_{11}$ has recently determined by Yamane's group.\cite{Yamane2014}
Similar to the structural character of
LaSi$_{3}$N$_{5}$, the crystal structure of
La$_{3}$Si$_{6}$N$_{11}$ also consists of corner-sharing
SiN$_{4}$ tetrahedra. La$_{3}$Si$_{6}$N$_{11}$
crystallizes in a tetrahedral structure with the space group of P4bm.  In this compound, there are
two non-equivalent La$^{3+}$ crystallographic sites. Both sites are coordinated with
eight nitrogen atoms, and we call them La$_{2a}$ and La$_{4c}$ according to their Wyckoff positions. The La$_{2a}$ site has a four-fold local symmetry,
 while the symmetry at the La$_{4c}$ site is lower (As seen in Figure 1). 

\begin{figure*}
        \centering
        \begin{subfigure}[b]{0.3\textwidth}
                \includegraphics[scale=0.2]{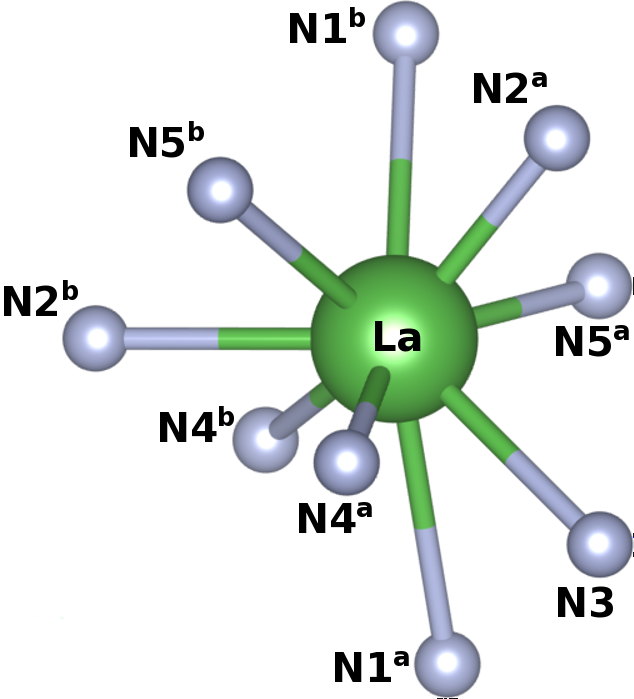}
                \caption{}
                \label{}
        \end{subfigure}%
          \quad   
        ~ 
        \begin{subfigure}[b]{0.3\textwidth}
                \includegraphics[scale=0.2]{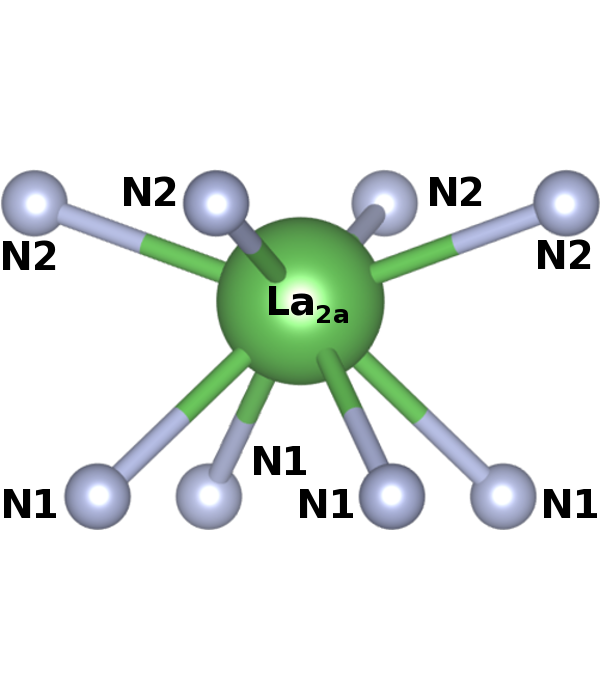}
                \caption{}
                \label{}
        \end{subfigure}
        \quad
        \begin{subfigure}[b]{0.3\textwidth}
                \includegraphics[scale=0.2]{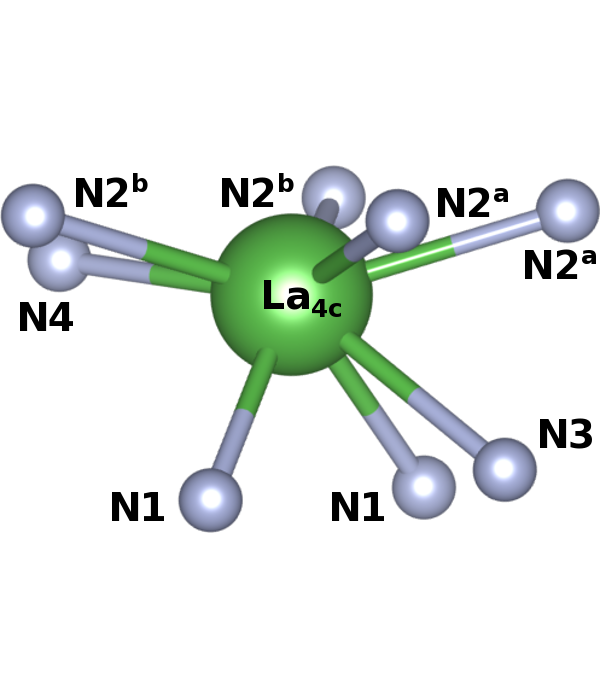}
                \caption{}
                \label{}
        \end{subfigure}%
            
\caption{ Coordination environment of (a) the La$^{3+}$ crystallographic site in
LaSi$_{3}$N$_{5}$; (b) the La$_{2a}$ site in
La$_{3}$Si$_{6}$N$_{11}$; (c) the La$_{4c}$ site in
La$_{3}$Si$_{6}$N$_{11}$. Green and gray spheres stand for La
and N atoms, respectively.}\label{figure_1}
\end{figure*}

In recent years, Ce$^{3+}$ doped LSN samples have been
synthesized to examine their potential in solid state lighting applications. The optical
performance of LaSi$_{3}$N$_{5}$:Ce was experimentally studied in detail.\cite{Suehiro2009,cai2009} These results show that the Ce$^{3+}$ ion
occupies the La$^{3+}$ crystallographic site, and gives the above-mentioned blue emission under UV excitation. 
On the other hand, the properties of
La$_{3}$Si$_{6}$N$_{11}$:Ce were investigated by Seto \textit{et al},\cite{Kijima2009} who found the yellow emission of
La$_{3}$Si$_{6}$N$_{11}$:Ce. This experimental observation has been recently confirmed by Seshadri's group.\cite{George2013} According to these facts, some preliminary discussions based on crystal structure and dielectric properties have been given to understand the difference between LaSi$_{3}$N$_{5}$:Ce and La$_{3}$Si$_{6}$N$_{11}$:Ce.\citep{Mikami2010,Mikami2013,George2013} However,there are still two main questions existing now, which
the semi-empirical method cannot answer: (1) Why do Ce$^{3+}$ ions emit different colors
 in the two LSN phosphors? (2) Which site is the luminescent center in the
La$_{3}$Si$_{6}$N$_{11}$:Ce phosphor?  

At present, first-principle calculations have been widely used in material science, as they
provide an useful insight in the chemical and electronic properties of materials and hence can aid in
the search for better materials or guide modification of existing
ones.\cite{Ponce2013,Samuel2013,Ponce2015,Marsman2000,Erhart2015,Erhart2014}
Compared to the semi-empirical model, they might also have advantages in the design of phosphors:
first, such theoretical simulations can provide a detailed understanding of interactions and effects involved
in the optical processes; second, without relying on empirical information
they have the potential to precisely simulate $4f\rightarrow 5d$ neutral transition in new potential materials, not only for the absorption process but also for the emission process.
Still, a routine use of first-principle methods might not yield the sought understanding. As an example, a recent first-principle study of LaSi$_{3}$N$_{5}$, doped with Ce as well as some other dopant ions, has been published.\cite{Ibrahim2014} 
The hybrid functional HSE06 has been used. They point out that, in their scheme, simply based on the Kohn-Sham band structure, the Ce$_{5d}$ states are in the conduction band, at variance with experimental data.

With this background, in the present study, we have performed an ab-initio study of these two LSN phosphors, aiming at answering the above-mentioned questions. Our methodological approach differs from the previous ab-initio study,\cite{Ibrahim2014} at three levels: (1) Constrained DFT allows us to obtain the Ce$_{5d}$ state below the conduction band, in agreement with experimental data; (2) ${\Delta}$SCF method is used to get the  $4f\rightarrow 5d$ transition energy of the Ce$^{3+}$ ion; (3) The lattice relaxation in the excited state is considered, yielding the Stokes shift and characterization of the
Ce$^{3+}$ luminescence. Our paper is structured as follows. In
Sec.\ref{num_approach}, we summarize the calculation method. The study of the undoped LSN bulk materials is
presented in Sec.\ref{bulk}. The
Ce$^{3+}$ doped calculations are shown in Sec.\ref{LSNCe}, and we give the conclusion in Sec.\ref{conclusion}.
\section{Numerical approach}
\label{num_approach}
\subsection{Computational details}
\label{comp_det}

In this work, the calculations were performed within density functional theory (DFT) using the
projector augmented wave (PAW) method as implemented in the ABINIT
package.\cite{Blochl1994,Abinit2002,Abinit2009,Marc2008} 
Exchange-correlation (XC) effects were treated within the
generalized gradient approximation (GGA).\cite{Perdew1996} 
For the Ce$^{3+}$
doped calculation, DFT+U was used, allowing the Ce$_{4f}$ states to be located inside the band gap.\cite{Liechtenstein1995} 
The U value has been
optimized to 4.6 eV (J = 0.5 eV), in order to reproduce the location of the $4f$ inside the gap,
following the hybrid functional study of LaSi$_{3}$N$_{5}$:Ce$^{3+}$.\cite{Ibrahim2014} 
Because of the similar composition of the two nitrides, the same U (J) value was also used in the study of
La$_{3}$Si$_{6}$N$_{11}$:Ce$^{3+}$.

Most of the PAW atomic datasets were directly taken from the ABINIT website.\cite{Jollet2014}
The nitrogen dataset, with $2s^{2}2p^{3}$ valence electrons, has a 1.3 Bohr radius
and two projectors per angular momentum channel. The silicon PAW dataset, with
$3s^{2}3p^{2}$ valence electrons, has a 1.71 Bohr sphere radius and two
projectors per angular momentum channel. The cerium dataset, with
$5s^{2}5p^{6}6s^{2}5d^{1}4f^{1}$ valence electrons, 
has a 2.5 Bohr radius and two projectors per angular momentum channel. 

For La, we tested two different PAW atomic datasets. The normal ABINIT
La PAW dataset, with $5s^{2}5p^{6}6s^{2}5d^{1}4f^{0}$
as valence electrons, has a 2.5 Bohr sphere radius and two projectors per angular channel. Here, we call
this La PAW atomic dataset `La$_{4f}$-semicore'. 
In order to test the effect of La$_{4f}$ orbitals, we have generated a PAW atomic dataset with the valence configuration $5s^{2}5p^{6}6s^{2}5d^{1}$ (freezing the unoccupied $4f$ orbital in the core). Here we denote this PAW atomic dataset as `La$_{4f}$-core'. 
The PAW atomic dataset was generated using the ATOMPAW software with the same input parameter as the
other La PAW GGA-PBE atomic dataset, but with frozen $4f$ states.

With these PAW atomic datasets, we performed the structural relaxation and band structure calculations. The
convergence criteria have been set to 10$^{-5}$ Hartree/Bohr (for residual forces) and 0.5 mHa/atom (for the tolerance on the total energy). In these calculations, cutoff kinetic energies of 30 Ha and 25 Ha for the plane-wave basis set were used for the 
LaSi$_{3}$N$_{5}$(:Ce) and La$_{3}$Si$_{6}$N$_{11}$(:Ce) compounds, respectively. The
Monkhorst-Pack sampling of the primitive cells (36 atoms for LaSi$_{3}$N$_{5}$ and 40 atoms for La$_{3}$Si$_{6}$N$_{11}$) for the same tolerance criteria were determined to be 3x3x2 and 3x3x3 for the two nitrides, respectively.

\subsection{Supercell calculations}
\label{supercell}

The Ce$^{3+}$ doped LSN calculations have been
conducted using the supercell method. The cluster approach based on Hartree-Fock theory was not
considered since previous studies have shown this method may lead to the deficiencies of the local
geometry around the rare earth site.\cite{Gracia2008,Pascual2008} 
Moreover, the obvious demerit of
the cluster method is the lack of the electronic information about the host lattice. We will see that the
position of the conduction band minimum (CBM) of the host is important in the luminescent center
identification. Therefore, a 72-atom 2x1x1 supercell for
LaSi$_{3}$N$_{5}$ and a 80-atom 1x1x2 supercell for
La$_{3}$Si$_{6}$N$_{11}$ have been used for the study of
Ce$^{3+}$ doped phosphor. In the calculations, one cerium atom substitutes one
lanthanum atom, which leads to La$_{7}$CeSi$_{24}$N$_{40}$ for
LaSi$_{3}$N$_{5}$:Ce, and two different cases,
La$_{11}$Ce$_{2a}$Si$_{24}$N$_{44}$ and
La$_{11}$Ce$_{4c}$Si$_{24}$N$_{44}$, for
La$_{3}$Si$_{6}$N$_{11}$:Ce. 
Such supercells have a Ce$^{3+}$ doping concentration of 12.5\% and 8.3\% for
the two LSN phosphors, respectively, which is reasonable compared to the experimental
data.\cite{Suehiro2009,Kijima2009}

\subsection{Neutral excitations}
\label{neutr_exc}

At present, the Bethe-Salpeter Equation (BSE)~\cite{Onida2002} 
is the best approach to study the optical properties of solids. It describes 
neutral excitations as coherent superpositions of electron-hole pairs.
However, the computational burden of such approach is quite heavy, 
and not feasible with supercells of nearly one hundred atoms.
Instead of BSE, we simulated the $4f\rightarrow 5d$ neutral excitation of Ce$^{3+}$ ion on the basis
of the constrained DFT method (CDFT). The electron-hole interaction, an essential contribution in the BSE, is mimicked by promoting the Ce$_{4f}$ electron to the
Ce$_{5d}$ state, by constraining the seven $4f$ bands to be unoccupied, while occupying the lowest $5d$ state lying higher in energy. The CDFT method has been used for the search of
novel scintillators, proposed by Canning, Chaudhry and coworkers.\cite{Canning2011,Chaudhry2014,Chaudhry2011} 
However, the previous
works aimed at the qualitative description of the RE$_{5d}$ state. Compared to
these results, we observe the CDFT ability to yield quantitative predictions
following the ${\Delta}$SCF method\cite{Tozera2000,Martin,van2014} that is, relying on total energy differences of the different constrained configurations.
Through the combination of CDFT and ${\Delta}$SCF, the $4f\rightarrow 5d$ neutral excitation of Ce$^{3+}$ ion is correctly described within the DFT framework, which has a much lower computational cost than the BSE method. In practice, the calculations are performed by setting manually the energy-ordered occupation numbers (optional independent input variables in ABINIT) to one or zero.
For both the ground state and the excited state, all valence levels, spin up as well as spin down, are occupied. For the ground state, additionally, the lowest spin-up level found in the band gap (clearly identified as a predominantly Ce-4f level) is occupied, while for the excited neutral state, the seven spin-up predominantly Ce-4f levels that are found in the band gap are unoccupied, and the next spin-up level is occupied. In most cases, the latter level is detached from the conduction band, and exhibits Ce-5d character, as will be seen later.

\subsection{Configurational coordinate diagram}
\label{conf_coord}

\begin{figure}
\includegraphics[scale=0.45]{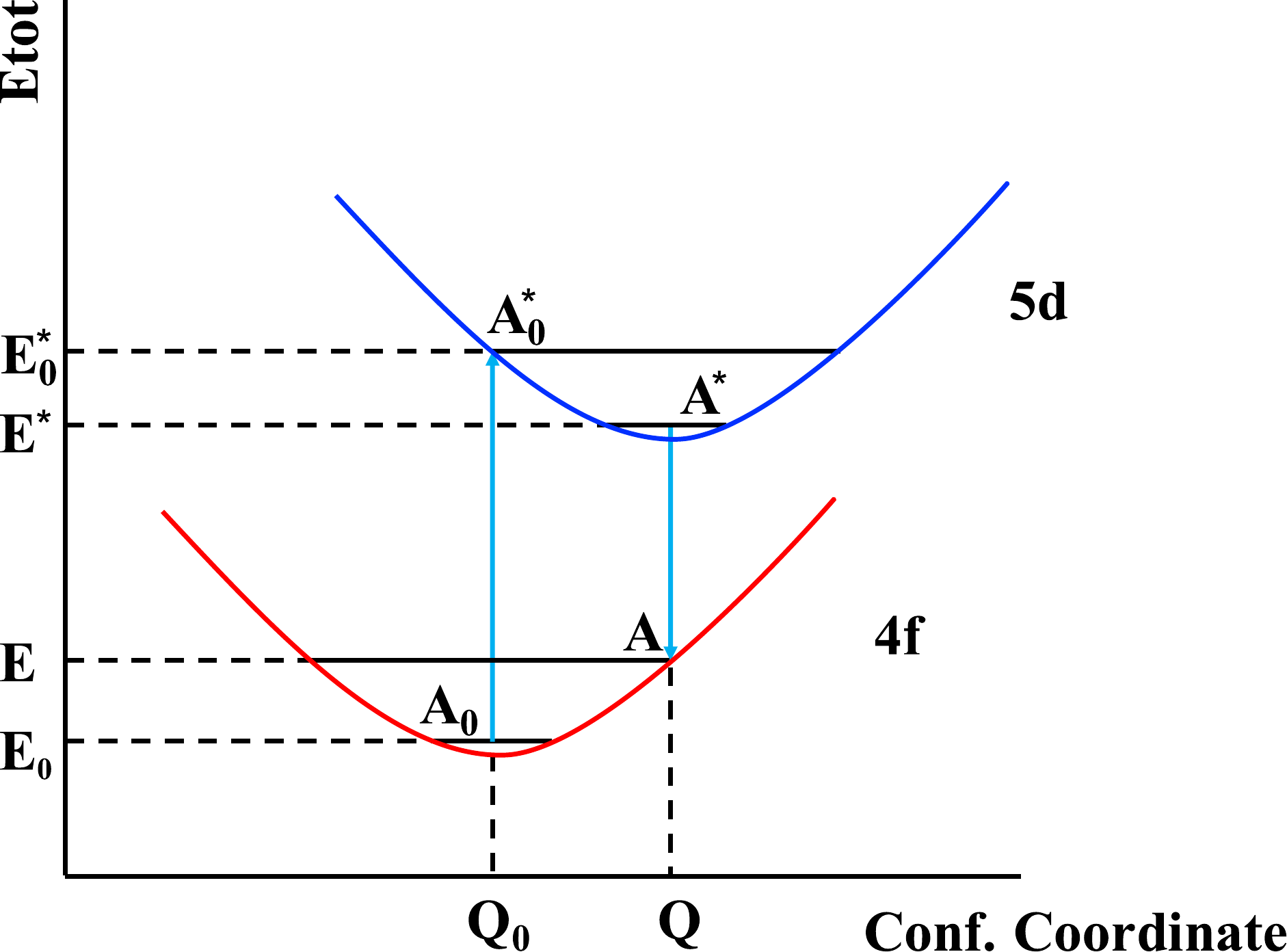}
\caption{Configurational coordinate diagram}\label{figure_2}
\end{figure}

Following the proposed method to describe the neutral excitation of Ce$^{3+}$ ion, we
will use the configuration coordinate diagram\cite{Blasse} to analyze the
absorption/emission process and Stokes shift of the two Ce$^{3+}$ doped lanthanum
silicate nitride phosphors, as shown in Figure \ref{figure_2}.

Here, we briefly explain its physical meaning. This configurational coordinate diagram depicts the
total energy of a system containing Ce$^{3+}$ion in its ground
state and of a system containing Ce$^{3+}$ion in its excited state, curve $4f$ and $5d$,
respectively, as a function of the generalized configuration coordinate Q, which can be made up of
any relevant combination of ionic degrees of freedom in the system. Q$_{0}$ and
Q represent the equilibrium configuration coordinates, for the system with
Ce$^{3+}$ in its ground state and in its excited state, respectively. The horizontal
lines inside the curves $4f$ and $5d$ denote the energy levels of the system in which the quantization of vibrational motion is taken into account. When a photon is absorbed by
the Ce$_{4f}$ electron, the Ce$^{3+}$ ion will be excited from the ground
state to the excited state, corresponding to A$_0\rightarrow$A$_0^{*}$. After the absorption, the system will be out of equilibrium due to the change of electronic configuration of the Ce$^{3+}$ion. The atomic positions are then relaxed following the forces, which is represented by the process A$_0^{*}\rightarrow$A$^*$ in Figure \ref{figure_2}.
After this lattice relaxation, the system reaches the new equilibrium state, at which the
emission process A$^*\rightarrow$A occurs. The cycle is finished by the lattice relaxation
A$\rightarrow$A$_0$ process in the ground state. Based on this idea, the absorption/emission energy and
the Stokes shift of Ce$^{3+}$doped phosphors can be determined as below:
\begin{eqnarray}
E_{abs} &=& E_{0}^{*}  - E_{0} \\
E_{em} &=& E^{*} - E\\
\Delta S &=& E_{abs } - E_{em } 
\end{eqnarray}

These values can be directly compared with experimental data, which can validate the proposed method
and yield the identification of the luminescence site.

\section{The pristine hosts}
\label{bulk}

In this section, we focus on the results for bulk LaSi$_{3}$N$_{5}$ and
La$_{3}$Si$_{6}$N$_{11}$: relaxed crystal structure, electronic band structure, and the role of the
La$_{4f}$ state in the calculation.

\subsection{Crystal structure}
\label{cryst_struct}

\begin{figure}
        \centering
        \begin{subfigure}[]{0.5\textwidth}
                \includegraphics[scale=0.35]{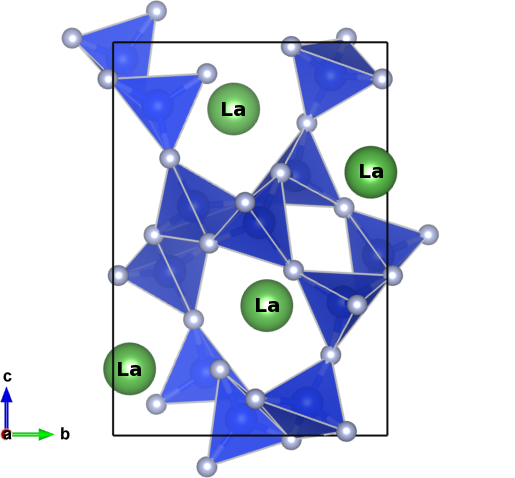}
                \caption{}
                \label{}
        \end{subfigure}%
          \quad  
          \\     
        ~ 
        \begin{subfigure}[]{0.5\textwidth}
                \includegraphics[scale=0.35]{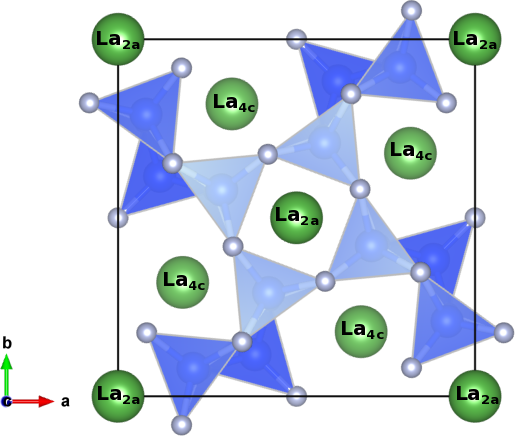}
                \caption{}
                \label{}
        \end{subfigure}
        
\caption{Crystal structure of (a) LaSi$_{3}$N$_{5}$, view
from \textit{a} direction; (b) La$_{3}$Si$_{6}$N$_{11}$, view from
\textit{c} direction. Green, blue and gray spheres stand for La, Si and N atoms, respectively.}\label{figure_3}
\end{figure}

\begin{table}
\centering
\renewcommand\arraystretch{1.5}
\caption{Lattice parameters of LSN bulk.}
\label{table_1}
\begin{tabular}{c|c|c|c}
\hline\hline
LSN & $a$ (\AA) & $b$ (\AA) & $c$ (\AA)
\tabularnewline
\hline
 LaSi$_{3}$N$_{5}$ calc. &
4.834 &
7.891 &
11.387\\
LaSi$_{3}$N$_{5}$ exp.\cite{Inoue1980} &
 4.807 &
 7.838 &
11.236\\
La$_{3}$Si$_{6}$N$_{11}$ calc. &
10.246 &
 10.246 &
 4.887\\
La$_{3}$Si$_{6}$N$_{11}$ exp.\cite{Yamane2014} &
10.199 &
10.199 &
4.841
\tabularnewline
\hline\hline
\end{tabular}
\end{table}

\begin{table}
\centering
\renewcommand\arraystretch{1.5}
\caption{Relaxed La-N bond lengths [\AA] of the La crystallographic site in
LaSi$_{3}$N$_{5}$ and La$_{3}$Si$_{6}$N$_{11}$.}
\label{table_2}
\begin{tabular}{cccc}
\hline\hline
Bond &
GGA-PAW&
Exp.\cite{Inoue1980,Yamane2014}\\
\hline
\multicolumn{3}{c}{LaSi$_{3}$N$_{5}$} \\
\hline
La-N1$^a$&
3.151&
3.134
\\
La-N1$^b$&
3.140&
3.067
\\
La-N2$^a$&
2.443&
2.421
\\
La-N2$^b$&
2.724&
2.697
\\
La-N3&
2.884&
2.832
\\
La-N4$^a$&
2.577&
2.553
\\
La-N4$^b$&
2.898&
2.884
\\
La-N5$^a$&
2.693&
2.645
\\
La-N5$^b$&
2.860&
2.830
\\
\hline
\multicolumn{3}{c}{La$_{3}$Si$_{6}$N$_{11}$ }\\
\hline
La$_{2a}$-N1(x4)&
2.652&
2.650
\\
La$_{2a}$-N2(x4)&
2.659&
2.644
\\
La$_{4c}$-N1(x2)&
2.528&
2.551
\\
La$_{4c}$-N2$^a$(x2)&
2.674&
2.674
\\
La$_{4c}$-N2$^b$(x2)&
2.893&
2.853
\\
La$_{4c}$-N3&
2.823&
2.863
\\
La$_{4c}$-N4&
2.640&
2.623
\tabularnewline
\hline\hline
\end{tabular}
\end{table}

In Section \ref{intro}, we have briefly mentioned that LaSi$_{3}$N$_{5}$
is built of SiN$_{4}$ tetrahedra, which are linked by shared corners. 
The La atom is centrally located between the pentagonal holes along
the c-axis, and is coordinated with nine nitrogen atoms at a distance between 2.6-3.2 \AA.
LaSi$_{3}$N$_{5}$ crystallizes in the orthorhombic crystal system with space group 
P2$_{1}$2$_{1}$2$_{1}$.

The space group of
La$_{3}$Si$_{6}$N$_{11}$ is more symmetric, P4bm. 
La$_{3}$Si$_{6}$N$_{11}$ is isostructural with
Ce$_{3}$Si$_{6}$N$_{11}$.
Similar to the structural character of
LaSi$_{3}$N$_{5}$, the crystal structure of
La$_{3}$Si$_{6}$N$_{11}$ also consists of corner-sharing
SiN$_{4}$ tetrahedra.
The experimental data~\cite{Inoue1980,Yamane2014} have been used to start the initial optimization of the crystal structure (with the La$_{4f}$-semicore PAW atomic dataset). Table \ref{table_1} lists the
relaxed lattice parameters for the two nitride compounds. 
The theoretical results for LSN bulk are consistent with the
experimental value, within 2\% relative difference. This small difference might be attributed to the
GGA exchange and correlation functional. 
Figure \ref{figure_3} shows the optimized crystal structure of the two nitrides. One can see 
that both compounds are
composed of corner-sharing SiN$_{4}$ tetrahedra, that form a  dense network
with large voids accommodating the La$^{3+}$ ions. 
Figure \ref{figure_1} depicts the relaxed coordinate environments of La$^{3+}$ ion in the two
nitrides: one La$^{3+}$ site in
LaSi$_{3}$N$_{5}$, and two non-equivalent La$^{3+}$ sites, La$_{2a}$
and La$_{4c}$ sites in La$_{3}$Si$_{6}$N$_{11}$. When doped, the
Ce$^{3+}$ ion is expected to substitute these crystallographic sites, which
results in one case of Ce$_{La}$ for LaSi$_{3}$N$_{5}$, and two cases
(Ce$_{2a}$ and Ce$_{4c}$) for
La$_{3}$Si$_{6}$N$_{11}$ with the Ce ion occupying the
La$_{2a}$ and La$_{4c}$ sites, respectively. 
Table \ref{table_2} lists the relaxed La-N bond lengths for the three La$^{3+}$ crystallographic
sites. Through these results, the average distance of La-N has been determined to be 2.819 {\AA} for the La site in the LaSi$_{3}$N$_{5}$, and 2.711/2.655 {\AA} for the La$_{2a}$/La$_{4c}$ site in the La$_{3}$Si$_{6}$N$_{11}$ compound. 

The nomenclature for the neighboring nitrogen atoms is presented in Figure \ref{figure_1} . This will help understand the different types of bonds presented in Table \ref{table_2} . In LaSi$_{3}$N$_{5}$, there are five types
of nitrogen atoms, while the lanthanum atom has nine neighbors. Some symmetry operations do not leave the lanthanum site unchanged. So, among the nine nearest-neighbor atoms of lanthanum, there are four pairs of equivalent nitrogen atoms (by symmetry), whose distance to the specific nitrogen atom differs. The local geometry of La$^{3+}$ site can be represented as a tricapped trigonal prism. In the more symmetric La$_{3}$Si$_{6}$N$_{11}$, there are four types of nitrogen atoms. The tetragonal symmetry leaves the La$_{2a}$ site unchanged, and the eight nearest neighbors nitrogen atoms of this site are split in two groups of four, where the La-N distances in each are equal. The La$_{4c}$ site is left unchanged by a mirror plane, so that five distances characterize the locations of the eight nearest-neighbour atoms. The local geometry of La$_{2a}$ and  La$_{4c}$ sites can be seen as a square antiprism and a bicapped trigonal prism, respectively. 

\subsection{Electronic band structure}
\label{elec_struct}

\begin{figure}
\begin{tabular}{cc}
\includegraphics[scale=0.5]{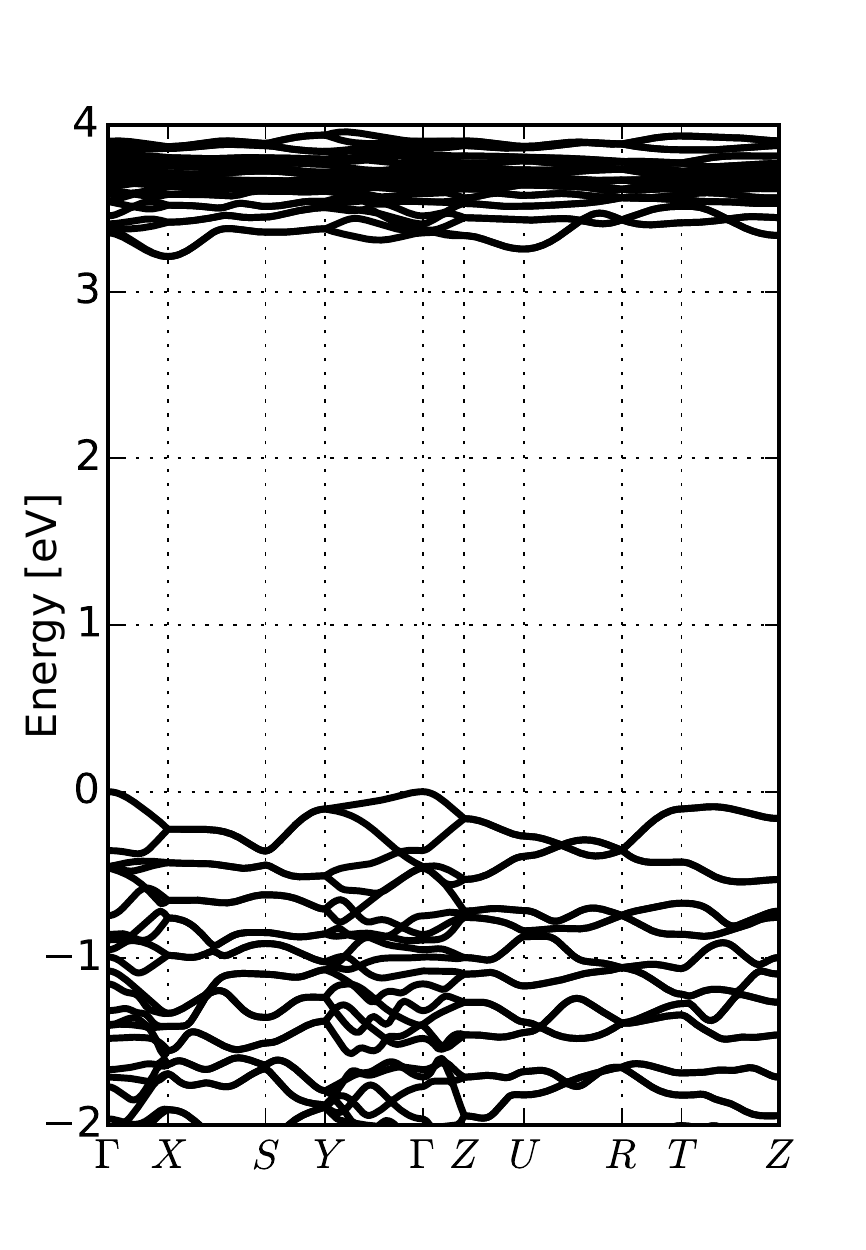} &
\includegraphics[scale=0.5]{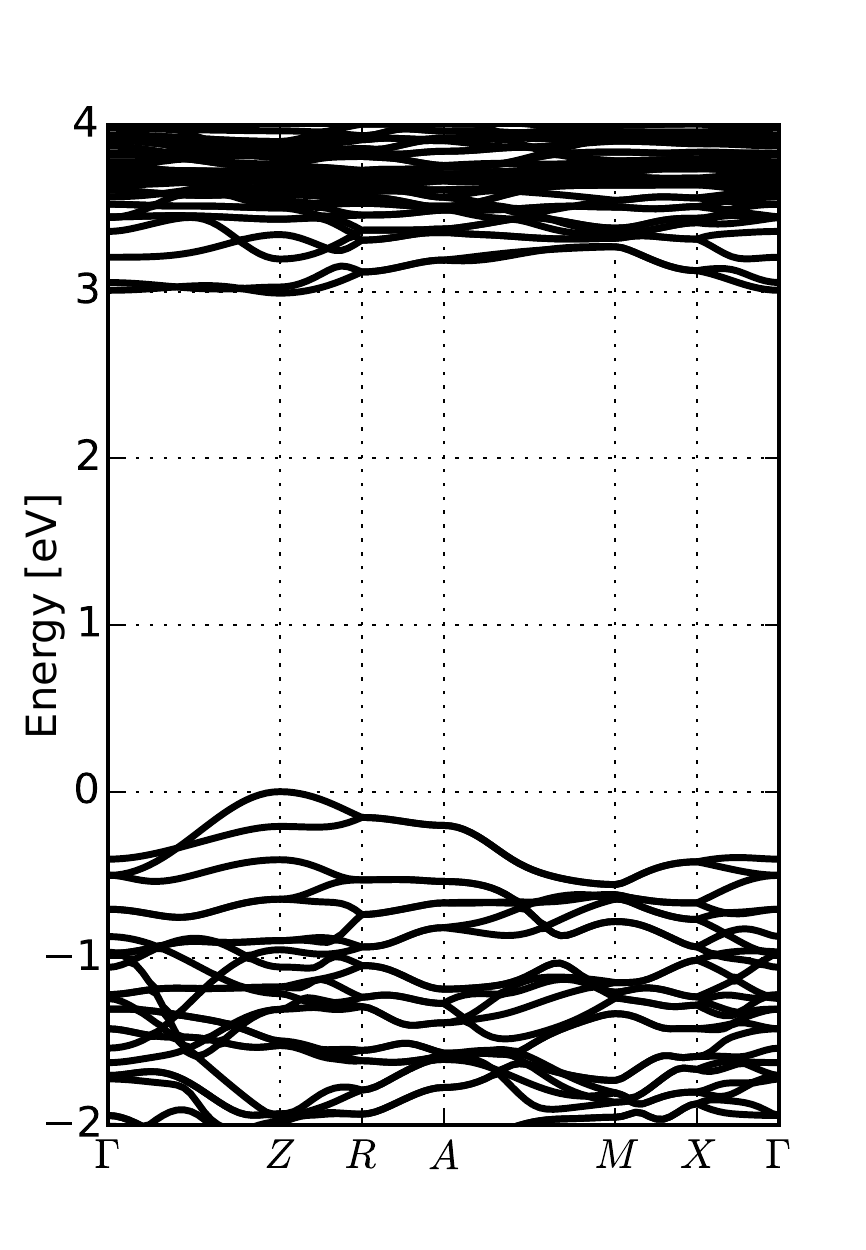} \\
\end{tabular}
\caption{ Kohn-Sham DFT electronic band structure of LSN bulk with converged lattice parameters.
LaSi$_{3}$N$_{5}$ (left)
La$_{3}$Si$_{6}$N$_{11}$ (right)}\label{figure_4}
\end{figure}

With the relaxed crystal structure of the two nitrides, we have calculated the corresponding
electronic band structure within the DFT framework, as it might have a bearing on the different luminescent
behaviors. Figure \ref{figure_4} shows the calculated results for the La$_{4f}$-semicore PAW atomic dataset. LaSi$_{3}$N$_{5}$ has a 3.21 eV $\Gamma-X$ indirect band gap, and La$_{3}$Si$_{6}$N$_{11}$ belongs to the
direct-transition class of compounds with a 2.99 eV band gap at the Z point. At present, there is no experimental data
for the band gap of the two compounds. However, we can expect that our
calculation suffers from the well-known `band gap problem' of DFT, with sizeable underestimation compared to experimental results. Recently, Ibrahim \textit{et al.} have
investigated the electronic structure of LaSi$_{3}$N$_{5}$ using the HSE06
functional, for which the band gap problem is much reduced compared to our GGA approach, giving a band gap of 4.8 eV.\cite{Ibrahim2014}
In Ref.\onlinecite{George2013}, a similar work for
La$_{3}$Si$_{6}$N$_{11}$ was conducted by Seshadri's group. A band gap
of about 4 eV was obtained for this compound.
Thus, the band gap seems consistently 
smaller in La$_{3}$Si$_{6}$N$_{11}$ than in LaSi$_{3}$N$_{5}$. A previous work of one of us was devoted to understand the origin of the change of band gaps, and tried to link it with the optical performance of Ce$^{3+}$ ion in the two compounds.\cite{Mikami2013} Here, we continue the study, aiming at providing a quantitative description about the performance of the two phosphors.  

\begin{figure}
        \centering
        \begin{subfigure}[]{0.5\textwidth}
                \includegraphics[scale=0.32]{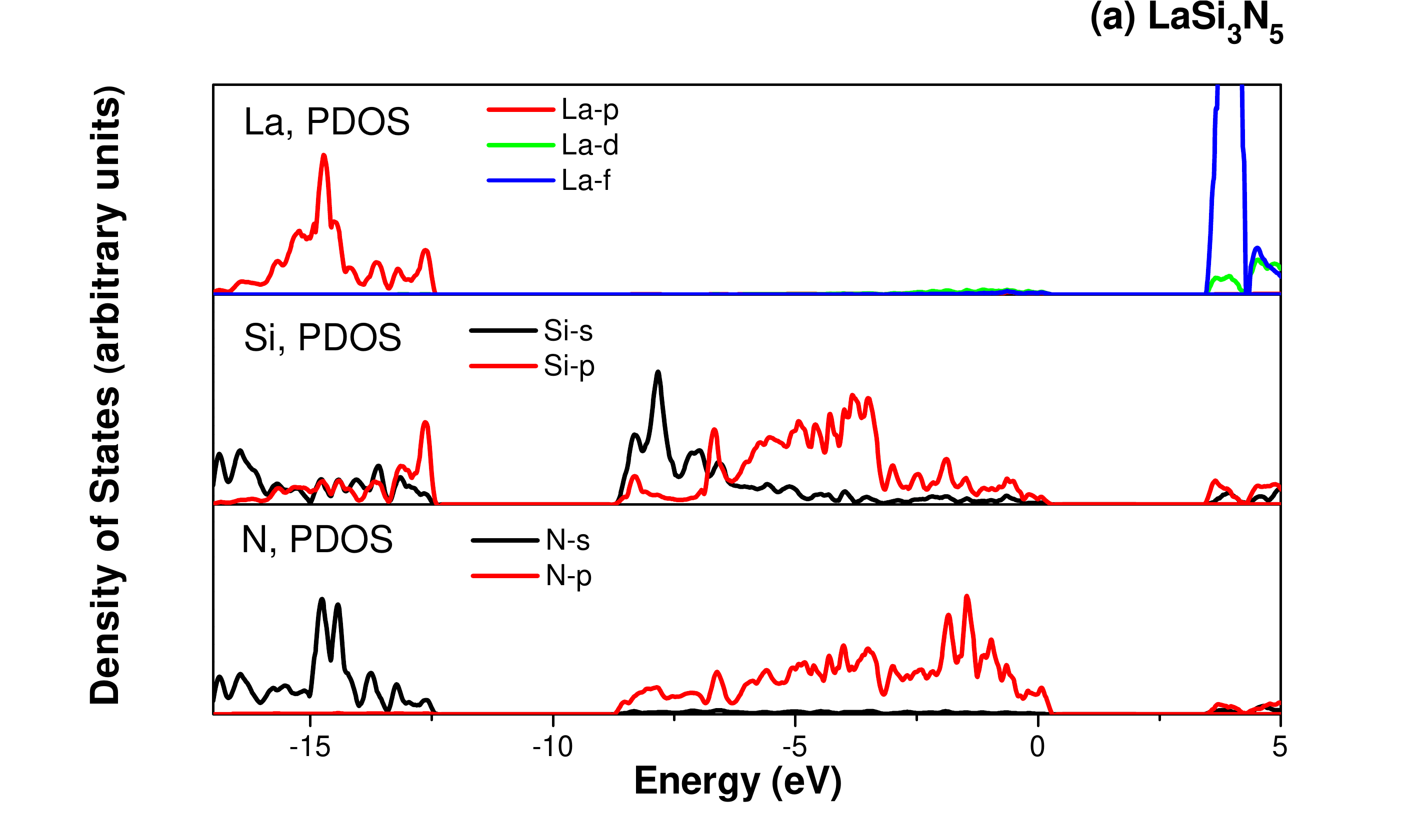}
        \end{subfigure}%
          \quad   
        ~ 
        \begin{subfigure}[]{0.5\textwidth}
                \includegraphics[scale=0.32]{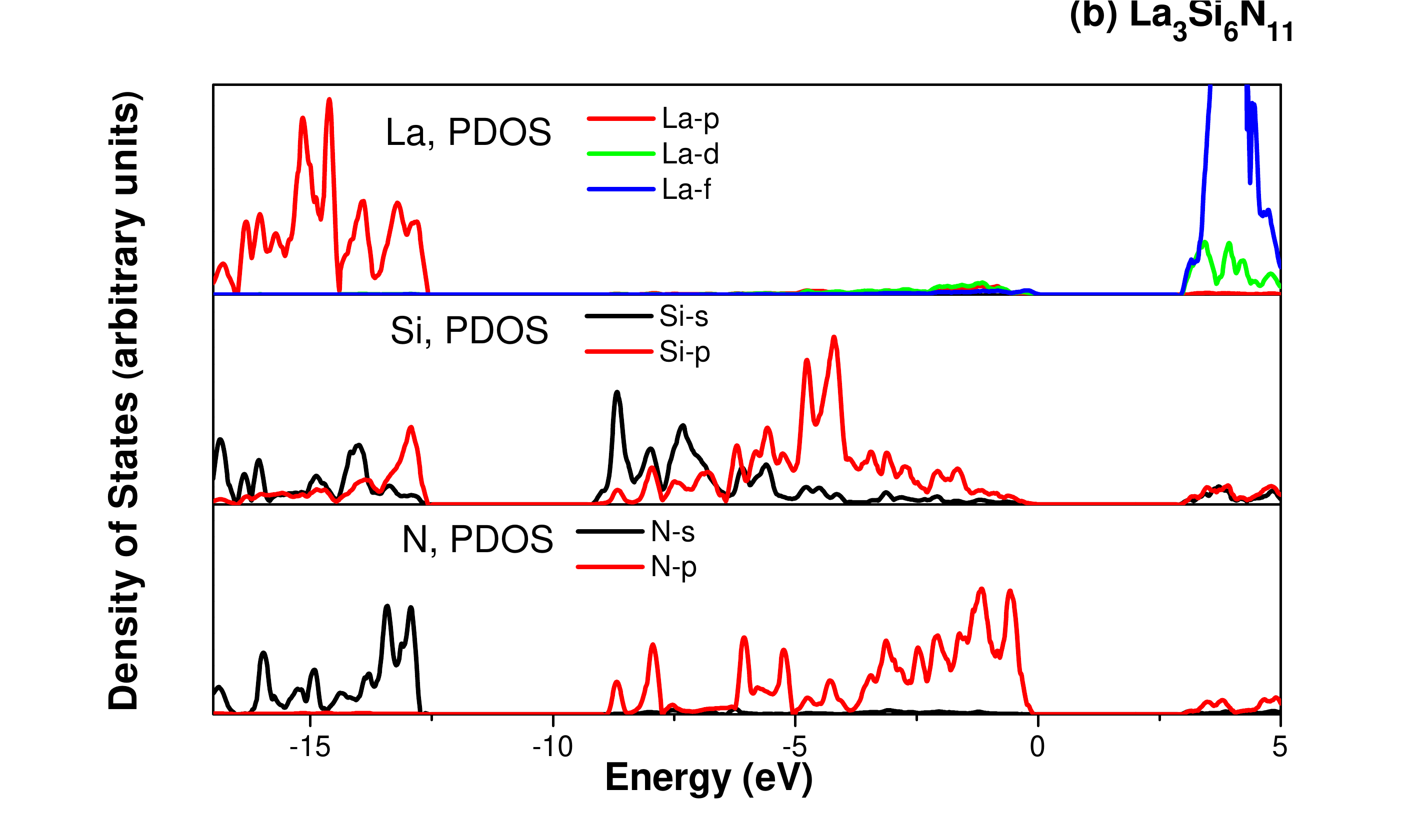}
        \end{subfigure}
        
\caption{Partial density of states of (a) LaSi$_{3}$N$_{5}$ and (b) La$_{3}$Si$_{6}$N$_{11}$}\label{figure_5}
\end{figure}

The composition of the valence band maximum (VBM) and
conduction band minimum (CBM) might also be important to understand the luminescence. We have 
examined the partial density of states, as shown in Figure \ref{figure_5}. The results show that for both compounds, the CBM consists of a mixture of La$_{4f}$ and La$_{5d}$ states, while the VBM mainly comes from N$_{2p}$ states. Such similarity again does not help to quantitatively explain the observed different optical
properties of LSN phosphors.

\begin{figure}
\centering
\begin{tabular}{cc}
\includegraphics[scale=0.5]{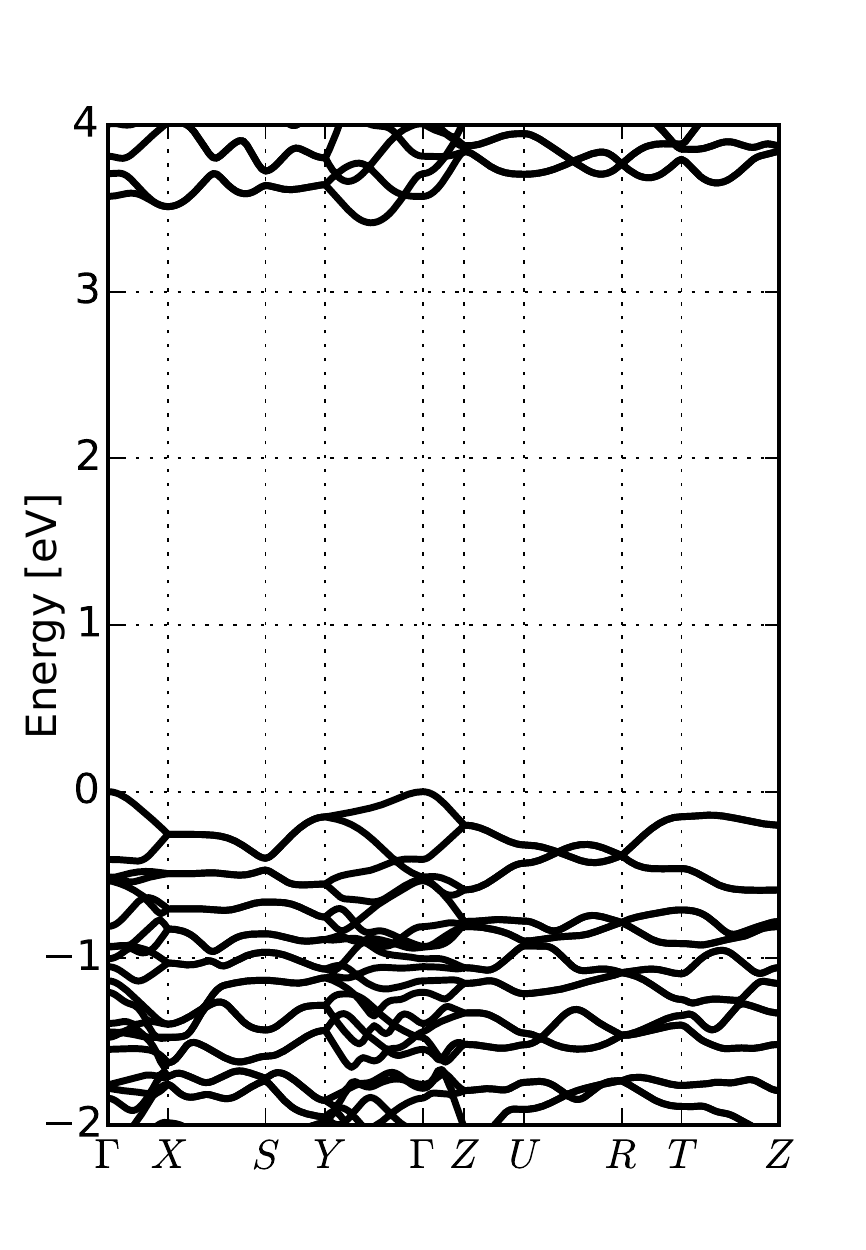} &
\includegraphics[scale=0.5]{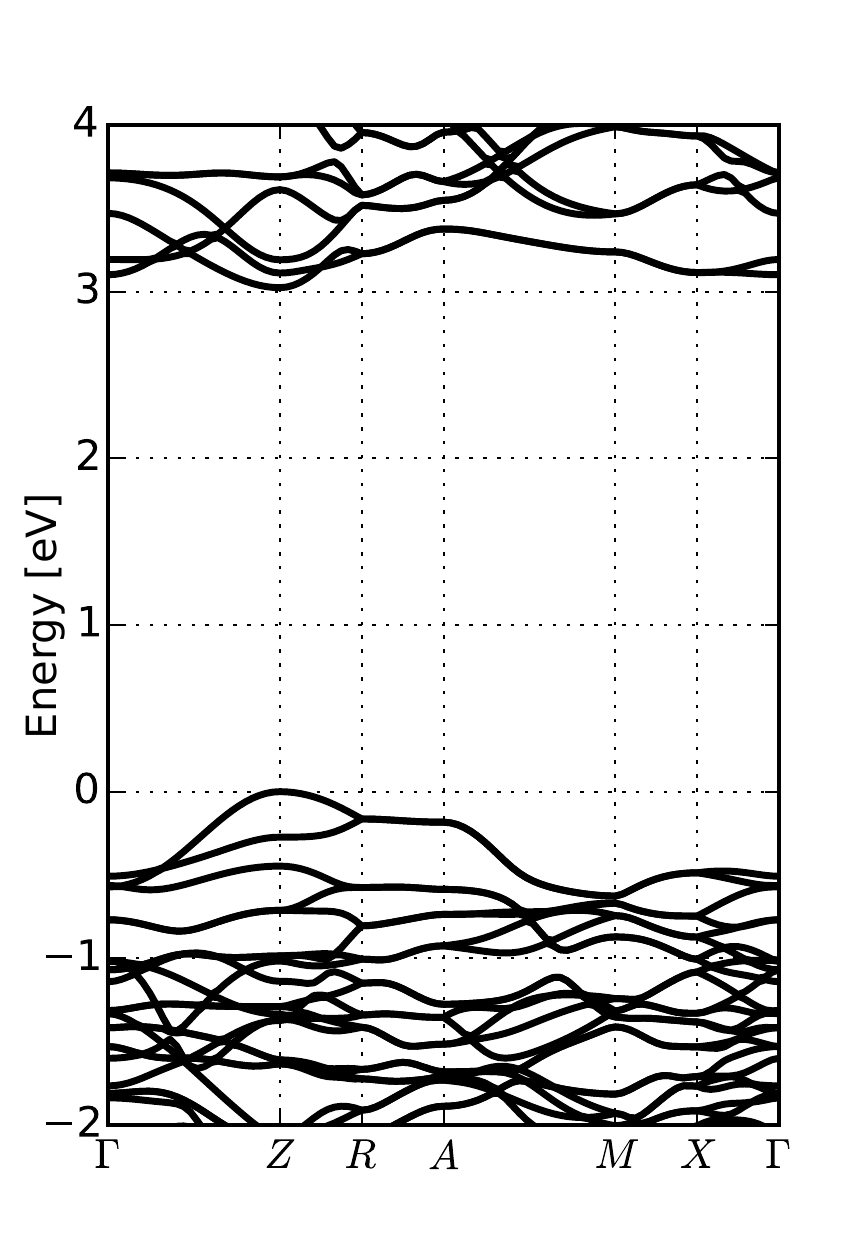}\\
\end{tabular}
\caption{Kohn-Sham DFT electronic band structure of LSN bulk, in which the 4f orbital contributions have been removed, thanks to the use of a PAW atomic dataset in which they are present in the core (La$_{4f}$-core). The structural properties are the same as in Figure 4 for LaSi$_{3}$N$_{5}$ (left) and La$_{3}$Si$_{6}$N$_{11}$ (right).}
\label{figure_6}
\end{figure}

To confirm the CBM composition of LSN bulk, the pseudopotential La$_{4f}$-core
has been used in band structure calculations. Figure \ref{figure_6} depicts the obtained band structure
with the La$_{4f}$-core PAW atomic dataset (still with the relaxed geometry from the lattice relaxation
based on La$_{4f}$-semicore PAW dataset). Through comparison of the Figures \ref{figure_4} and \ref{figure_6}, it can be seen that
the $4f$ freezing does not modify the valence band, but clearly impacts the conduction band because
of the missing $4f$ state in the La$_{4f}$-core case. The modification of the conduction
band leads to small changes of band gap, the one of LaSi$_{3}$N$_{5}$ becomes 3.34 eV and
the one of La$_{3}$Si$_{6}$N$_{11}$ becomes 2.91 eV. From these results, it can
be deduced that the La$_{4f}$ state should be considered into our ab-initio calculation,
in order to provide accurate results for the CBM and CBM-related states.

\section{Cerium-doped materials}
\label{LSNCe}

The above bulk study has provided the basic information for the two LSN compounds. 
 In this section, the results of Ce$^{3+}$
doped calculations will be presented, including the ground state of the supercell, the excited state description, the effect of lattice relaxation in the excited state and the luminescent center identification.
The supercells used in this part, namely
La$_{7}$CeSi$_{24}$N$_{40}$,
La$_{11}$Ce$_{2a}$Si$_{24}$N$_{44}$ and
La$_{11}$Ce$_{4c}$Si$_{24}$N$_{44}$, will be
simply denoted as LaSi$_{3}$N$_{5}$:Ce,
La$_{3}$Si$_{6}$N$_{11}$:Ce$_{2a}$ and
La$_{3}$Si$_{6}$N$_{11}$:Ce$_{4c}$. 

\subsection{Ground state}
\label{ground_state}

As mentioned in Section \ref{num_approach}, the analysis of the Ce$^{3+}$ luminescence will be based on the
configurational coordinate diagram. Therefore, we first focus our attention on the ground state
properties. 
Table \ref{table_3} lists the relaxed lattice parameters of the supercells. The data for the 
corresponding LSN bulk supercells is also
shown for comparison, which is obtained from the calculation results for the LSN primitive cell but
doubled in the \textit{a} direction for LaSi$_{3}$N$_{5}$, and in the \textit{c} direction
for La$_{3}$Si$_{6}$N$_{11}$. From these results, we see that
the Ce$^{3+}$ doping leads to a slight shrinkage of the crystal cell. The reason
for the contraction can be ascribed to the relatively smaller ionic radius of Ce$^{3+}$
ion than the one of La$^{3+}$ ion
. The ionic radius of Ce$^{3+}$ is determined to 1.143 {\AA} and 1.196 {\AA} for the coordination number of 8 and 9, respectively, while the corresponding values for the La$^{3+}$ ion are 1.16 {\AA}  and 1.216 {\AA} for the La$^{3+}$ ion.\cite{Shannon1969} 

With the relaxed geometry obtained above, the ground state electronic band structures for
LaSi$_{3}$N$_{5}$:Ce,
La$_{3}$Si$_{6}$N$_{11}$:Ce$_{2a}$ and
La$_{3}$Si$_{6}$N$_{11}$:Ce$_{4c}$, have been
calculated. The corresponding total energies are listed in Table \ref{table_4}. The values of La$_{3}$Si$_{6}$N$_{11}$:Ce$_{2a}$ and La$_{3}$Si$_{6}$N$_{11}$:Ce$_{4c}$ are quite similar, with La$_{3}$Si$_{6}$N$_{11}$:Ce$_{2a}$ being favored only by 0.3 mHa (less than 10 meV), which indicates that the Ce$^{3+}$ ions should nearly equally occupy the La$_{2a}$ and La$_{4c}$ sites in La$_{3}$Si$_{6}$N$_{11}$ at firing temperature(1500 - $\SI{2000}{\degreeCelsius}$). Moreover, there are twice more asymmetric sites than symmetric sites in La$_{3}$Si$_{6}$N$_{11}$. Hence, there should be twice more Ce$^{3+}$ lying in La$_{4c}$ sites than in  La$_{2a}$. This result is consistent with experimental observation that shows the Ce$^{3+}$ can equally substitute on the La$_{2a}$ and La$_{4c}$ site at high Ce concentration.\cite{George2013}

\begin{table}
\centering
\renewcommand\arraystretch{1.5}
\caption{Relaxed lattice parameters [\AA] and Volume [\AA$^3$] of LSN:Ce supercell.}
\label{table_3}
\begin{tabular}{c|c|c|c|c}
\hline\hline
LSN & Volume (\AA$^3$) & $a$ (\AA) & $b$ (\AA) & $c$ (\AA)
\tabularnewline
\hline
 LaSi$_{3}$N$_{5}$ &
868.77 &
9.668 &
7.891 &
11.387\\
LaSi$_{3}$N$_{5}$:Ce &
867.32 &
9.662 &
7.890 &
11.376\\
La$_{3}$Si$_{6}$N$_{11}$ &
1026.16 &
10.246 &
10.246 &
9.774\\
La$_{3}$Si$_{6}$N$_{11}$:Ce$_{2a}$ &
1025.78 &
10.244 &
10.244 &
9.774\\
La$_{3}$Si$_{6}$N$_{11}$:Ce$_{4c}$ &
1025.37 &
10.243 &
10.243 &
9.772
\tabularnewline
\hline\hline
\end{tabular}
\end{table}

\begin{figure*}
        \centering
\captionsetup[subfigure]{labelformat=empty}
        \begin{subfigure}[ground state case]{0.45\textwidth}
                \includegraphics[scale=0.37]{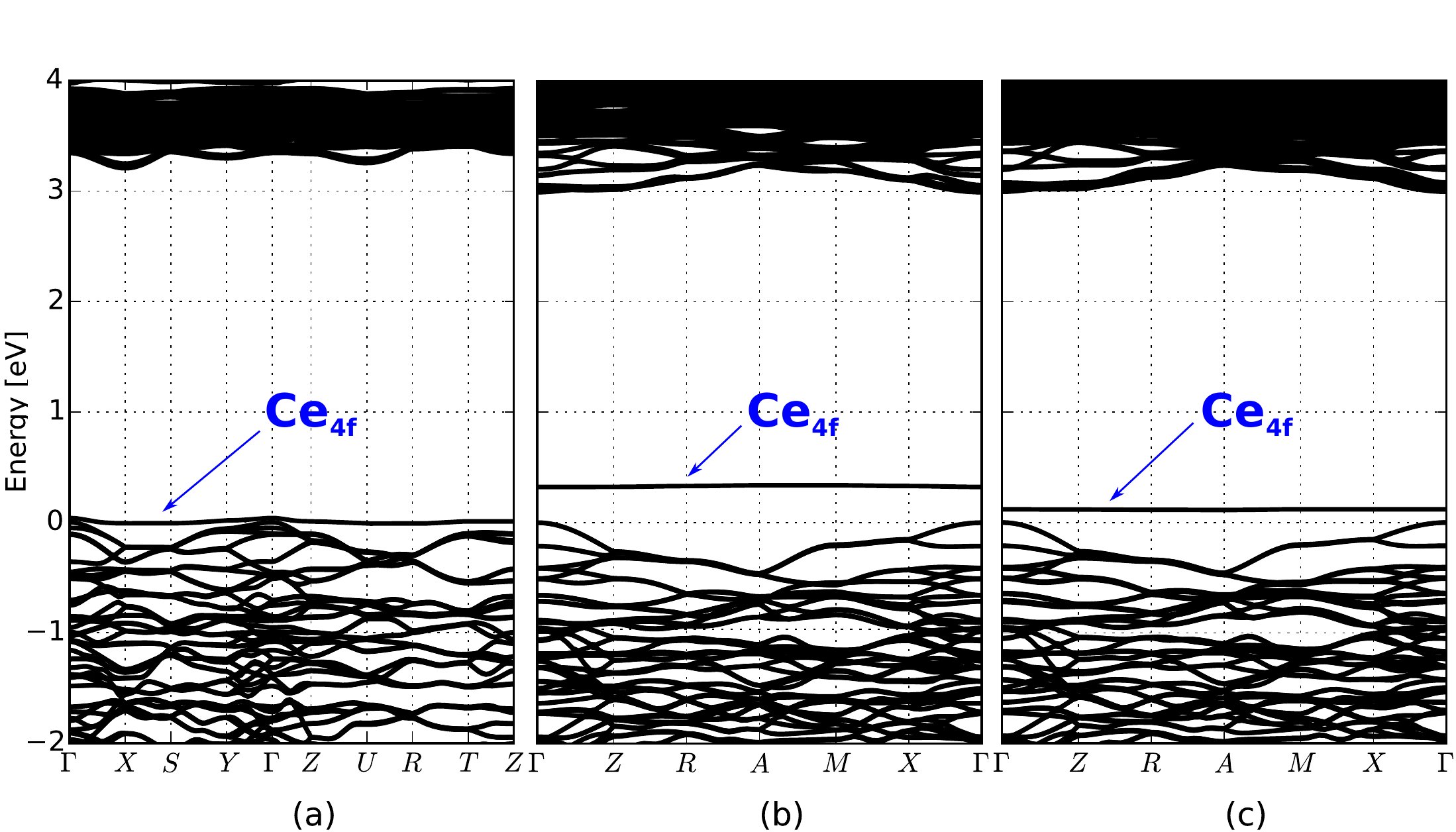}
                \caption{A$_0$ case, VB and CB}
                \label{}
        \end{subfigure}%
          \quad   
        ~ 
        \begin{subfigure}[]{0.45\textwidth}
                \includegraphics[scale=0.37]{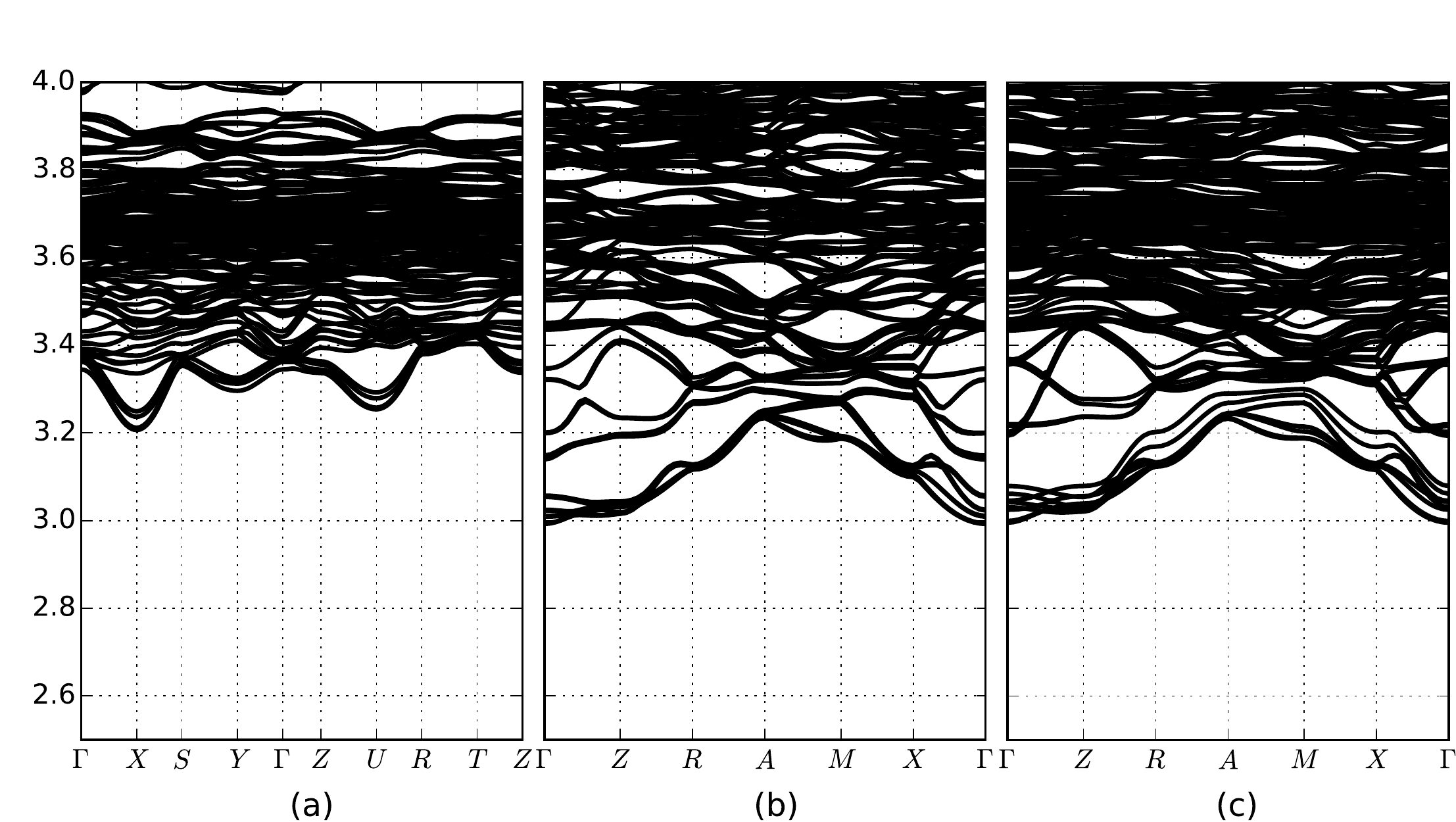}
                \caption{A$_0$ case, CB}
                \label{}
        \end{subfigure}
           \quad  
\captionsetup[subfigure]{labelformat=empty}
        \begin{subfigure}[]{0.45\textwidth}
                \includegraphics[scale=0.37]{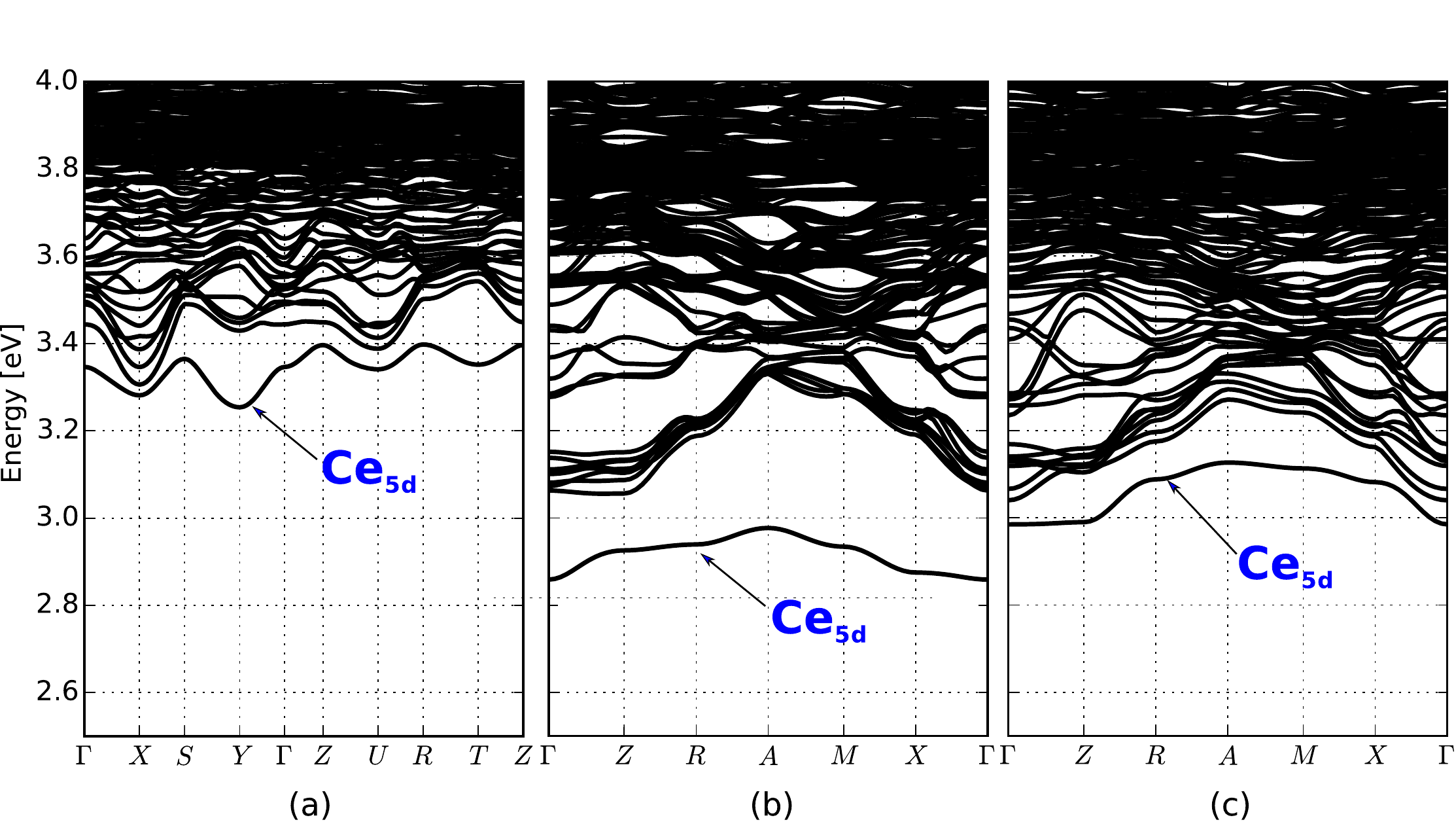}
                \caption{A$_0^*$ case}
                \label{}
        \end{subfigure}       
          \quad
        \begin{subfigure}[]{0.45\textwidth}
                \includegraphics[scale=0.37]{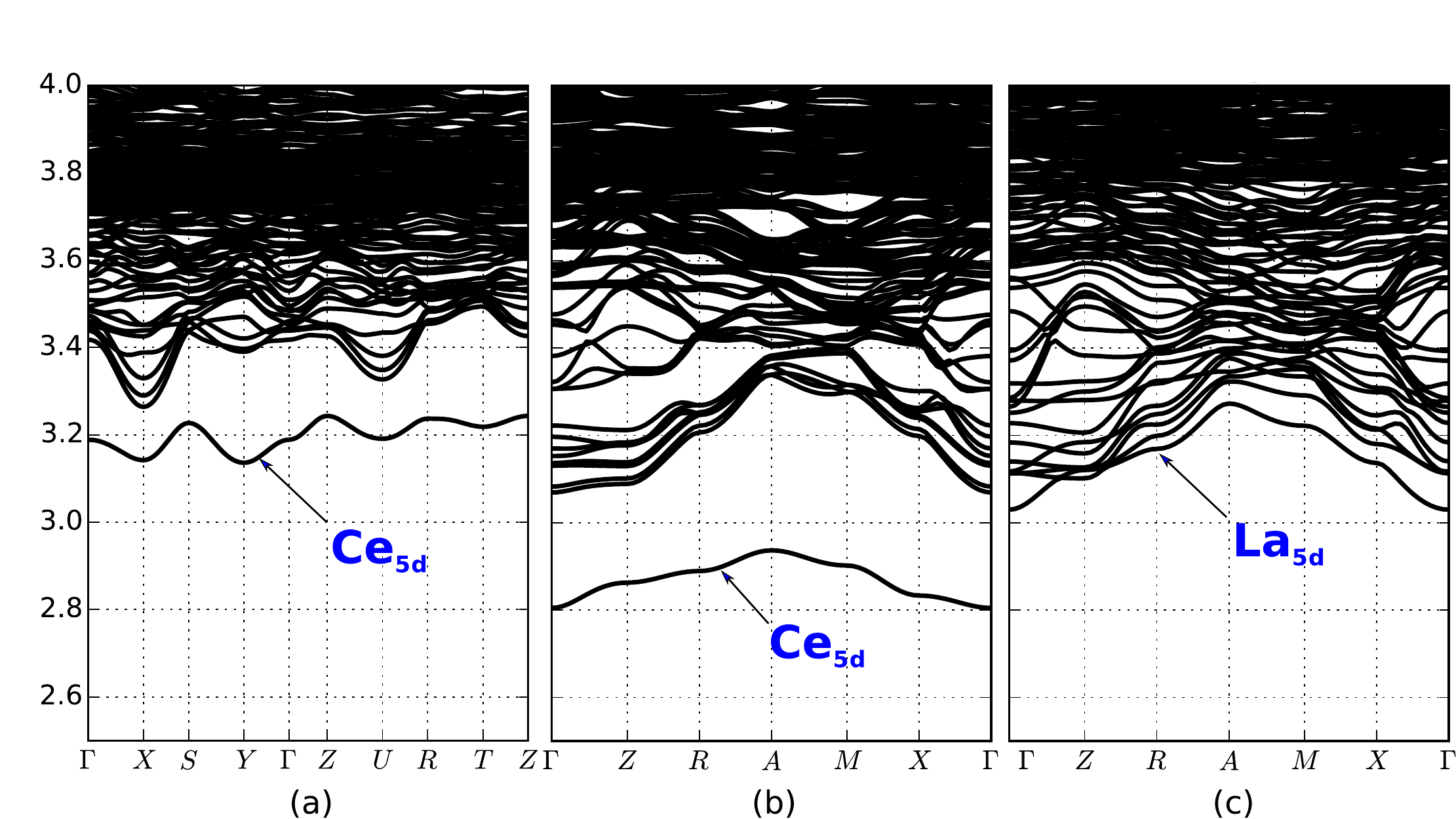}
                \caption{A$^*$ case}
                \label{}
        \end{subfigure}  
                  
\caption{Electronic band structure of: ground state A$_0$ case, valence band (VB) and conduction band (CB); ground state A$_0$ case, CB; excited state A$_0^*$ case, with ground state geometry; excited state A$^*$ case, with excited state geometry.(a) LaSi$_{3}$N$_{5}$:Ce; (b) La$_{3}$Si$_{6}$N$_{11}$:Ce$_{2a}$; (c) La$_{3}$Si$_{6}$N$_{11}$:Ce$_{4c}$.} 
\label{figure_7}
\end{figure*}

\begin{figure*}
\centering   
        \begin{subfigure}[b]{1\textwidth}
                \includegraphics[width=\linewidth]{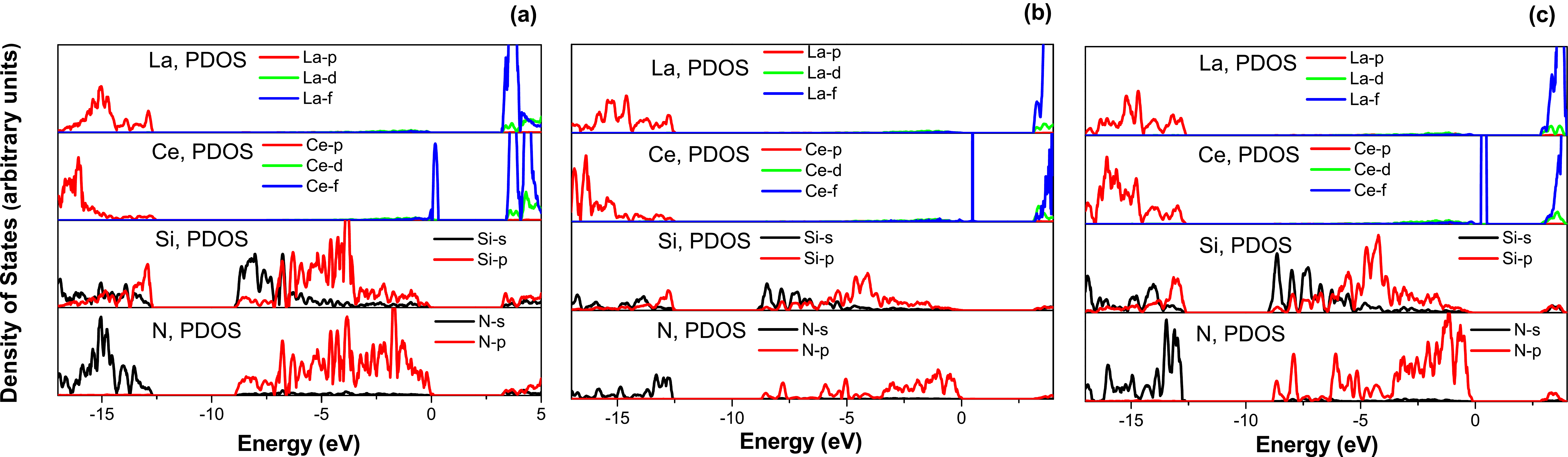}
        \end{subfigure}
        \quad
        \begin{subfigure}[b]{1\textwidth}
                \includegraphics[width=\linewidth]{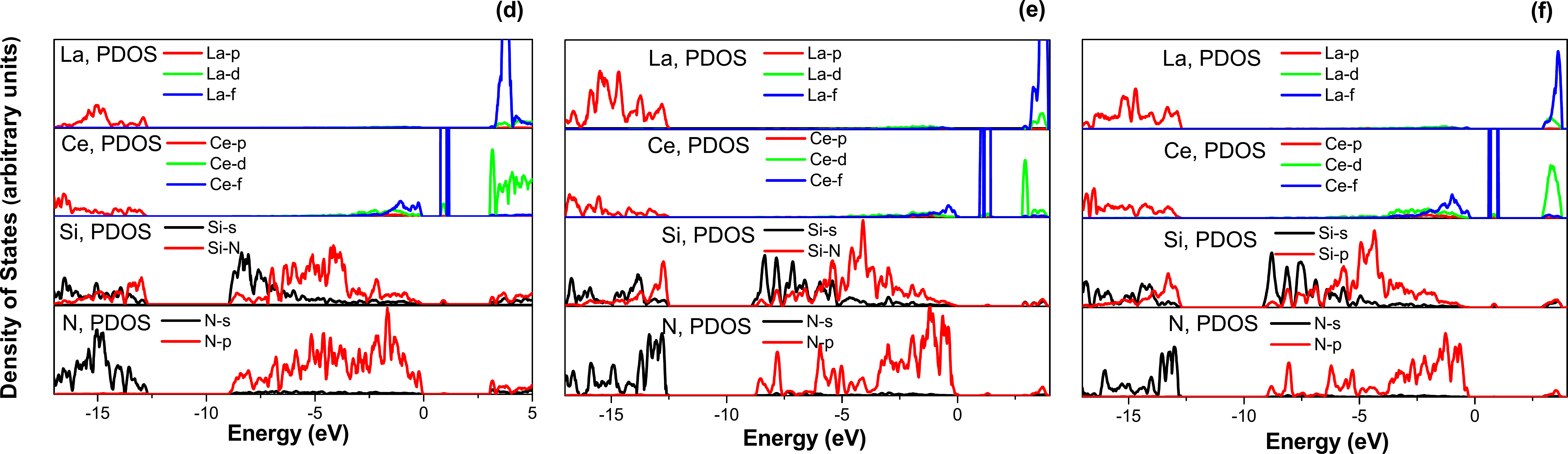}             
        \end{subfigure}      
\caption{Partial density of state of LSN:Ce phosphors: (a) LaSi$_{3}$N$_{5}$:Ce,A$_0$ case; (b) La$_{3}$Si$_{6}$N$_{11}$:Ce$_{2a}$,A$_0$ case; (c) La$_{3}$Si$_{6}$N$_{11}$:Ce$_{4c}$,A$_0$ case; (d) LaSi$_{3}$N$_{5}$:Ce,A$^{*}$ case; (e) La$_{3}$Si$_{6}$N$_{11}$:Ce$_{2a}$,A$^{*}$ case; (f) La$_{3}$Si$_{6}$N$_{11}$:Ce$_{4c}$,A$^{*}$ case.}
\label{figure_8}
\end{figure*}

Figure \ref{figure_7} (Left-hand, upside) shows the Kohn-Sham DFT band structure results, in the ground state (supercell). Compared with the bulk results shown in Figure \ref{figure_4}, for the primitive cell, a localized Ce$_{4f}$ state occurs in the band gap. 
In this case, no Ce$_{5d}$ state 
appears into the band gap, as indicated in Figure \ref{figure_7} (Right-hand, upside). Thanks to partial density of states plotting as shown in Figure \ref{figure_8}(a)-(c), the composition of CBM for the LSN
phosphors has been determined to be a hybrid state of La$_{4f}$ and
La$_{5d}$, similar to the result of the corresponding LSN bulk. This result
is not consistent with the experimental fact that both LSN phosphors give an intense emitting
light, for which a localized La$_{5d}$ state, inside the band gap, is expected. 
The reason for this failure is an incorrect identification of the levels in a ground-state electronic band structure, provided by DFT (or even quasiparticle band structure) with the experimentally observed neutral excitation of the system.
 Indeed, during the absorption process, the Ce$_{4f}$ electron is promoted to the
Ce$_{5d}$ state and a hole is left. Due to the localized nature of the Ce$_{4f}$
state, there is a strong Coulomb attraction between the promoted Ce$_{5d}$ electron
and the hole in the Ce$_{4f}$ state. This electron-hole interaction will pull the
Ce$_{5d}$ state to a lower energy, which is not depicted in a standard ground-state DFT
band structure.

\subsection{Excited state}
\label{exc_state}

As shown by the ground-state calculations, the electron-hole interaction is
necessary to describe the excited state of the LSN phosphor. In our work, the electron-hole
interaction is described in the constrained DFT method. Figure \ref{figure_7} (A$_0^*$ case) depicts the results for LSN phosphors. When one hole is created in the $4f$ band, the Ce$_{5d}$ state
successfully appears below the CBM of the LSN host. Eigenenergies from
the DFT calculations cannot be identified as the optical transition levels for the neutral
excitation. Instead of comparing eigenenergies, the neutral excitation energy can be
calculated from the ${\Delta}$SCF method through the total energy difference of the ground state and excited state. The calculated absorption energies for
LaSi$_{3}$N$_{5}$:Ce,
La$_{3}$Si$_{6}$N$_{11}$:Ce$_{2a}$ and
La$_{3}$Si$_{6}$N$_{11}$:Ce$_{4c}$ are shown in Table \ref{table_4}, and will be used to identify the luminescent center in the two LSN phosphors.

\subsection{Lattice relaxation in the excited state}
\label{lattice_relax_exc_state}

\begin{table}
\centering
\renewcommand\arraystretch{1.5}
\caption{The calculated absorption, emission and Stokes shift of LSN phosphors, compared with the experimental data.\citep{Suehiro2009,Kijima2009}}
\label{table_4}
\begin{tabular}{cccc}
\hline\hline
Case &
LaSi$_{3}$N$_{5}$:Ce &
La$_{3}$Si$_{6}$N$_{11}$:Ce$_{2a}$&
La$_{3}$Si$_{6}$N$_{11}$:Ce$_{4c}$ \\
\hline
A$_0$&
-763.4608 Ha&
-932.1088 Ha&
-932.1091 Ha
\\
A$_0^{*}$&
-763.3323 Ha&
-932.0062 Ha&
-932.9887 Ha
\\
A$^{*}$&
-763.3379 Ha& 
-932.0129 Ha&
-932.0924 Ha
\\
A&
-763.4524 Ha&
-932.1011 Ha&
-931.9968 Ha
\\
\hline
$\Delta$E$_{abs}$(A$_0^{*}$-A$_0$)&
3.50 eV&
2.79 eV&
3.28 eV
\\
$\Delta$E$_{abs}$(Exp.)&
3.43 eV&
2.58 eV&
-
\\
$\Delta$E$_{em}$(A$^{*}$-A)&
3.12 eV&
2.40 eV&
2.60 eV
\\
$\Delta$E$_{em}$(Exp.)&
2.95 eV&
2.25 eV&
-
\\
\hline
$\Delta$S(Cal.)&
3080 cm$^{-1}$&
3160 cm$^{-1}$&
5456 cm$^{-1}$
\\
$\Delta$S(Exp.)&
3815 cm$^{-1}$&
2717 cm$^{-1}$&
-
\\
\hline\hline
\end{tabular}
\end{table}

\begin{figure*}
        \centering
        \begin{subfigure}[b]{0.28\textwidth}
                \includegraphics[width=\linewidth]{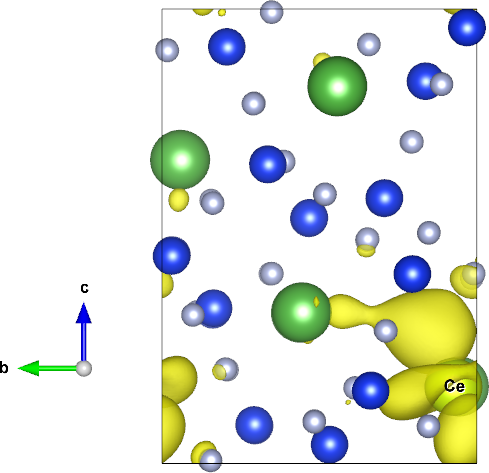}
                \caption{LaSi$_{3}$N$_{5}$:Ce}
                \label{}
        \end{subfigure}%
          \quad   
        ~ 
        \begin{subfigure}[b]{0.31\textwidth}
                \includegraphics[width=\linewidth]{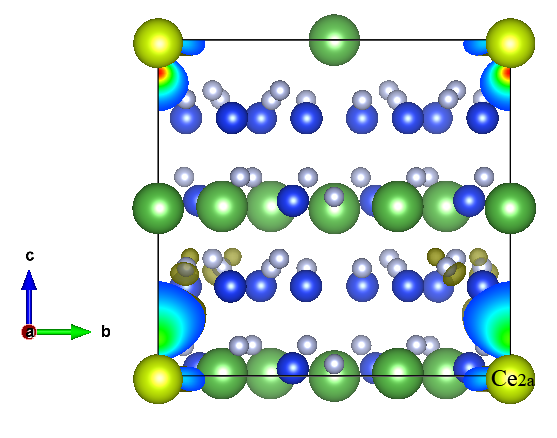}
                \caption{La$_{3}$Si$_{6}$N$_{11}$:Ce$_{2a}$}
                \label{}
        \end{subfigure}
        \quad
        \begin{subfigure}[b]{0.31\textwidth}
                \includegraphics[width=\linewidth]{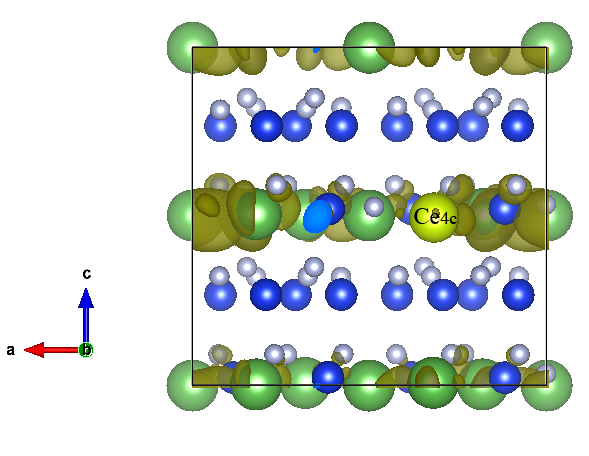}
                \caption{La$_{3}$Si$_{6}$N$_{11}$:Ce$_{4c}$}
                \label{}
        \end{subfigure}%
            
\caption{Charge density at $\Gamma$ point of (a) the 200$^{th}$ band of LaSi$_{3}$N$_{5}$:Ce (yellow isosurface); (b) the 232$^{nd}$ band of La$_{3}$Si$_{6}$N$_{11}$:Ce$_{2a}$ (blue and red isosurface, and some small yellow isosurface); (c) the 232$^{nd}$ band of La$_{3}$Si$_{6}$N$_{11}$:Ce$_{4c}$ (yellow isosurface).}
\label{figure_9}
\end{figure*}

Following the configuration coordinate diagram in Figure \ref{figure_2}, after absorption,
the system will be out of equilibrium due to the change in electronic configuration,
leading to the relaxation of atomic position in the excited state. Accordingly, we have conducted geometry optimization of doped LSN phosphors. The constrained DFT has been used to keep the electronic
configuration of Ce$^{3+}$ ion in its excited state. In the geometry optimization, the
lattice parameters have been fixed to the value from the ground state, because the time scale of atomic position relaxation is much shorter than the change of the macroscopic state (strain) of the
crystal.\cite{Blasse,Ponce2015}  
Figure \ref{figure_7} (A$^*$ case) shows the band structure for the excited state of emission process, and the corresponding partial density of states are depicted in the Figure \ref{figure_8}(d)-(f). Also, the charge density at the $\Gamma$ point of the lowest band over the Ce$_{4f}$ state is shown in Figure \ref{figure_9}. 
For the case of La$_{3}$Si$_{6}$N$_{11}$:Ce$_{4c}$, the lowest conduction band has La$_{5d}$ character, while those of LaSi$_{3}$N$_{5}$:Ce and La$_{3}$Si$_{6}$N$_{11}$:Ce$_{2a}$ are composed of Ce$_{5d}$ state.
Based on the above calculation, we have obtained the theoretical emission energies and
Stokes shifts for the LaSi$_{3}$N$_{5}$:Ce and
La$_{3}$Si$_{6}$N$_{11}$:Ce, that are listed in
Table \ref{table_4}. Through the comparison with the experimental value for the LSN phosphor, we can deduce that the
luminescence center for the LaSi$_{3}$N$_{5}$:Ce and
La$_{3}$Si$_{6}$N$_{11}$:Ce phosphors is from the
Ce$_{La}$ and Ce$_{2a}$ site, respectively. For the case of
La$_{3}$Si$_{6}$N$_{11}$:Ce$_{4c}$, we tentatively assign this Ce site as non-luminescent center with the Ce$_{5d}$ states inside the conduction bands, as we discussed in our work for other phosphors.\cite{Ponce2015} Although more advanced methods like the GW approximation might be needed to precisely determine the relative position of
Ce$_{5d}$ and the CBM, this is out of the scope of this work.\cite{Aryasetiawan1998} Here, it is worthy to note that the calculated Stokes shift for LaSi$_{3}$N$_{5}$:Ce is underestimated compared to the experimental data, while the calculated value for La$_{3}$Si$_{6}$N$_{11}$:Ce is overestimated. The reason for the above difference is due to the relative accuracy of the theoretical method, which gives an error about 0.2eV for the transition energies, while the accurate value of Stokes shift is smaller than 0.5eV. 

We observe that, with such excited structure
geometry, the Stokes shift and emission energy are in reasonable agreement with experimental data. 
At this stage, the effect of the excited state relaxation of atomic positions can be analyzed.
One can focus on the change of Ce$^{3+}$ coordination environment in the
LaSi$_{3}$N$_{5}$ and  La$_{3}$Si$_{6}$N$_{11}$, as listed in Table \ref{table_5}. The change of Ce-N bond length is
due to the movement of ions in the crystal structure. 
For LaSi$_{3}$N$_{5}$, an anisotropic distortion occurs around the Ce$^{3+}$ ion. Four bond lengths are increased (by up to 6\% for Ce-N1$^b$) while five bond lengths are decreased (by up to 5\% for Ce-N2$^b$ and Ce-N5$^b$). For La$_{3}$Si$_{6}$N$_{11}$:Ce$_{2a}$ there is a slight shortening of the Ce-N1 bond length, and a larger shortening of the Ce-N2 bond length (3\%). Finally, for La$_{3}$Si$_{6}$N$_{11}$:Ce$_{4c}$, all bond lengths contract, by up to nearly 7\% for Ce-N4. These information can be useful for the analysis of emission color of Ce$^{3+}$ ion in the two LSN compounds.

Now, we can come back to the analysis of the behaviour of the eigenenergy difference between the Ce$_{5d}$
band and the bottom of the conduction band. We will show that it can increase as well as decrease upon relaxation, depending on the behaviour of the 
energy difference between the Ce$_{4f}$ band and the bottom of the conduction band, and the bonding or anti-bonding character of the Ce$_{5d}$ orbital.
For this purpose, we first define the total energy $E(\textbf{R},f_{5d})$ as a function of the Ce$_{5d}$ occupation number, denoted $f_{5d}$, and at atomic positions symbolically denoted as $\bf R$. Note that the occupation of the Ce$_{4f}$ levels
decrease at the same rate as the occupation of the Ce$_{5d}$ increases.
Due to Janak's theorem, \cite{Janak1978} the derivative of the total energy with respect to the Ce$_{5d}$ occupation number is directly linked to the difference between Ce$_{5d}$ and Ce$_{4f}$ eigenvalues, as well as to their differences with respect to the conduction band minimum, 
$\frac {\partial {E(\textbf{R},f_{5d})}} {\partial f_{5d}}=\epsilon_{5d}-\epsilon_{4f}=(\epsilon_{5d}-\epsilon_{CBM})-(\epsilon_{4f}-\epsilon_{CBM})$.
In the exact density-functional theory, the eigenenergies themselves should not depend on the occupation numbers, while in semi-local approximations,
a convexity is observed.\cite{Coco2005} Actually, this convexity is quite important, and we have argued in the previous subsection that the difference of total energies is more reliable than eigenenergies to predict excited states and their relaxation. 
Still, if the eigenenergies were constant, 
$E(\textbf{R},f_{5d}=1)-E(\textbf{R},f_{5d}=0)=(\epsilon_{5d}-\epsilon_{CBM})-(\epsilon_{4f}-\epsilon_{CBM})$. Comparing the relaxed and unrelaxed geometries, we see that the Stokes shift is approximately equal to the change of $\epsilon_{5d}-\epsilon_{CBM}$ minus the change $\epsilon_{4f}-\epsilon_{CBM}$ upon relaxation. Thus, if the Ce$_{4f}$ eigenenergy increases significantly with respect to the conduction band, due to the relaxation, and in particular, if this increase is bigger than the Stokes shift,
it might be that the Ce$_{5d}$ raises with respect to the conduction band. We have checked that this is indeed the case for La$_{3}$Si$_{6}$N$_{11}$:Ce$_{4c}$. Physically, we know that the electronic negative charge that is present on the Ce$_{4f}$ in the ground state repels the negatively charge N ions. When the Ce$_{4f}$ gets unoccupied, a large contraction of the N cage is observed, as described in the previous subsection. If the Ce$_{5d}$ is more anti-bonding than the La$_{5d}$ states at the bottom of the conduction band, the eigenenergy difference between them will decrease.

\begin{table}
\centering
\renewcommand\arraystretch{1.5}
\caption{Ce-N bond lengths [\AA] of the Ce site.The bold characters highlight the biggest change of local geometry of three Ce$^{3+}$ sites, as mentioned in the text.}
\label{table_5}
\begin{tabular}{ccccc}
\hline\hline
\multicolumn{4}{c}{LaSi$_{3}$N$_{5}$:Ce}
\tabularnewline
\hline
Bond & Ground & Excited & Geometry \\
Ce-N1$^a$&
3.151&
3.178&
\multirow{9}{*}{\includegraphics[scale=0.16]{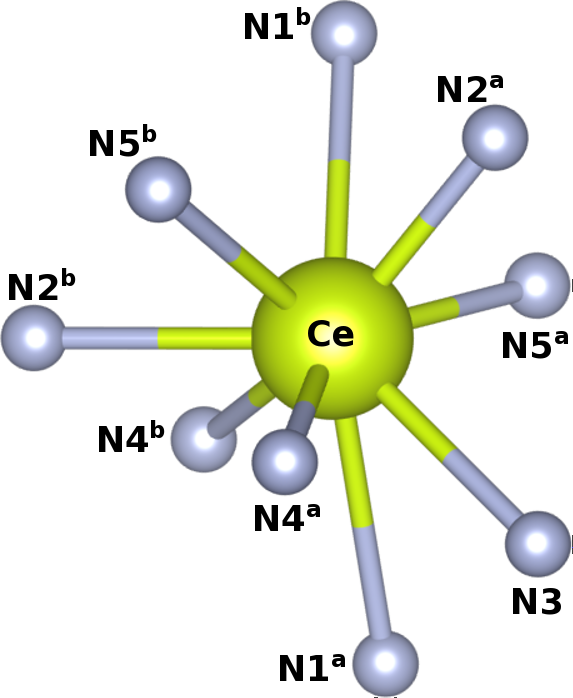}}
\\
Ce-N1$^b$&
3.137&
\textbf{3.324}&
\\
Ce-N2$^a$&
2.430&
2.380&
\\
Ce-N2$^b$&
2.717&
\textbf{2.601}&
\\
Ce-N3&
2.874&
2.955&
\\
Ce-N4$^a$&
2.553&
2.416&
\\
Ce-N4$^b$&
2.898&
2.835&
\\
Ce-N5$^a$&
2.690&
2.792&
\\
Ce-N5$^b$&
2.850&
\textbf{2.712}&
\\
\hline
\multicolumn{4}{c}{La$_{3}$Si$_{6}$N$_{11}$:Ce$_{2a}$}
\tabularnewline
\hline
\\
Ce-N1 (x4)&
2.657&
2.645&
\multirow{2}{*}{\includegraphics[scale=0.16]{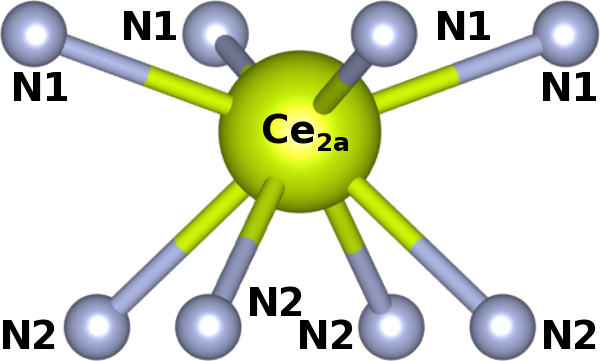}}
\\
\\
Ce-N2 (x4)&
2.638&
\textbf{2.555}&
\\
\\
\hline
\multicolumn{4}{c}{La$_{3}$Si$_{6}$N$_{11}$:Ce$_{4c}$}
\tabularnewline
\hline
\\
Ce-N1(x2)&
2.512&
2.389&
\multirow{5}{*}{\includegraphics[scale=0.16]{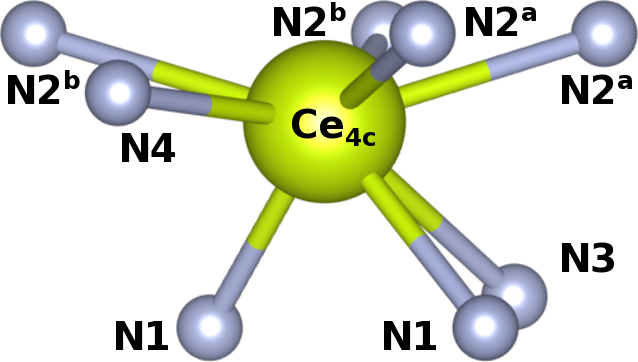}}
\\
Ce-N2$^a$ (x2)&
2.670&
2.596&
\\
Ce-N2$^b$ (x2)&
2.901&
2.895
\\
Ce-N3&
2.802&
2.718&
\\
Ce-N4&
2.641&
\textbf{2.472}&
\\
\hline\hline
\end{tabular}
\end{table}

\subsection{Analysis based on the Dorenbos model with first-principles geometry data}
\label{Dorenbos}

The Dorenbos model described in the the Appendix has been used to analyse the different luminescent behaviour of LaSi$_{3}$N$_{5}$:Ce and
La$_{3}$Si$_{6}$N$_{11}$:Ce. The structural geometries of the ground and excited state, as depicted in Table \ref{table_5}, were adopted to obtain the R$_{i}$ and R$_{av}$ in the corresponding formula. The results about the centroid shift of the 5d energy, $\varepsilon_{c}$, are listed in Table \ref{table_6}. At present, the $\varepsilon_{c}$ analysis of La$_{3}$Si$_{6}$N$_{11}$:Ce$_{4c}$ in the emission state was not conducted, since this site was determined to be non-luminescent. For the cases of LaSi$_{3}$N$_{5}$:Ce and
La$_{3}$Si$_{6}$N$_{11}$:Ce$_{2a}$, the obtained results indicate that the difference in $\varepsilon_{c}$ is 0.47eV and 0.5eV, for the ground and excited state, respectively. 

\begin{table}
\centering
\renewcommand\arraystretch{1.5}
\caption{Dorenbos model analysis of the [Xe]5d state of Ce$^{3+}$ ion in LSN phosphors. GS: ground state; EX: excited state}
\label{table_6}
\begin{tabular}{c|c|c|c}
\hline\hline
-- & LaSi$_{3}$N$_{5}$:Ce & La$_{3}$Si$_{6}$N$_{11}$:Ce$_{2a}$ & La$_{3}$Si$_{6}$N$_{11}$:Ce$_{4c}$
\tabularnewline
\hline
 $\chi_{av}$ &
1.74 &
1.68 &
1.68\\
\hline
$\alpha_{sp}^N$ &
 7.07 &
7.52 & 
7.52\\
\hline
$\varepsilon_{c}$, GS&
21380 cm$^{-1}$ &
25166 cm$^{-1}$ &
23671 cm$^{-1}$\\
\hline
$\varepsilon_{c}$, EX&
23950 cm$^{-1}$ &
28242 cm$^{-1}$ &
--\\
\hline\hline
 $\beta$&
5.67$\times$10$^8$ &
1.20$\times$10$^9$ &
1.20$\times$10$^9$\\
\hline
 R$_{av}$, GS&
 281 pm &
265 pm &
270 pm\\
\hline
 R$_{av}$, EX&
 280 pm &
260 pm &
--\\
\hline
$\varepsilon_{cfs}$, GS&
7181 cm$^{-1}$ &
17088 cm$^{-1}$ &
16461 cm$^{-1}$\\
\hline
$\varepsilon_{cfs}$, EX&
7232 cm$^{-1}$ &
17751 cm$^{-1}$ &
--\\
\hline\hline
\end{tabular}
\end{table}

Qualitative analysis also has been performed on the crystal field splitting, $\varepsilon_{cfs}$. In our cases, the coordination environment of Ce$^{3+}$ ion in LaSi$_{3}$N$_{5}$:Ce is in the form of tricapped trigonal prism, while those of Ce$_{2a}$ and Ce$_{4c}$ in La$_{3}$Si$_{6}$N$_{11}$:Ce can be seen as distorted cubic (square antiprism and bicapped trigonal prism, respectively). Following this idea, the $\varepsilon_{cfs}$ in the two nitrides can be calculated through the $\beta$, fitted according to Ref.\onlinecite{Dorenbos2000b}. The corresponding results are also listed in Table \ref{table_6}. Similar with the situation of $\varepsilon_{c}$, the analysis of La$_{3}$Si$_{6}$N$_{11}$:Ce$_{4c}$ in the emission state was not conducted. If we assume r(LSN) to be equal to 2.4, the effect of $\varepsilon_{cfs}$ on the cases of LaSi$_{3}$N$_{5}$:Ce and
La$_{3}$Si$_{6}$N$_{11}$:Ce$_{2a}$, give an energy difference of 0.51eV and 0.53eV, for the ground and excited state, respectively.

Through the analyses above, the difference of red-shift between the LaSi$_{3}$N$_{5}$:Ce and La$_{3}$Si$_{6}$N$_{11}$:Ce$_{2a}$, is calculated to be 0.98eV and 1.03eV for the ground and excited state that compares reasonably with the experimental data of 0.81eV and 0.72eV, respectively. The calculated Stokes shift is 2600cm$^{-1}$ and 3466cm$^{-1}$ for the LaSi$_{3}$N$_{5}$:Ce and La$_{3}$Si$_{6}$N$_{11}$:Ce$_{2a}$. Considering the qualitative character of Dorenbos model, these values are quite reasonable. Based on the above consistence, we can give the conclusion in two aspects: firstly, the structural geometry for LSN phosphor are accurately described, not only for the ground state, but also for the excited state. Second, the difference of luminescence in LaSi$_{3}$N$_{5}$:Ce and La$_{3}$Si$_{6}$N$_{11}$:Ce$_{2a}$ could be ascribed to the larger spectroscopic polarization of N$^{3-}$ ion in La$_{3}$Si$_{6}$N$_{11}$,\citep{Mikami2010,Mikami2013} and the stronger crystal field splitting of Ce$_{2a}$ site.

\section{Conclusion}
\label{conclusion}

In this paper, an ab-initio study has been conducted to accurately describe the neutral excitation
of Ce$^{3+}$ ions in two nitrides, LaSi$_{3}$N$_{5}$ and
La$_{3}$Si$_{6}$N$_{11}$. The analysis of Ce$^{3+}$
luminescence follows from the configurational coordinate diagram, in which the ground state and excited
state descriptions rely on the DFT+U and the constrained DFT approach. The absorption and emission
energies are calculated with the ${\Delta}$SCF method. Following these methods, the Stokes shift can be
obtained. The luminescent centers (Ce$_{La}$ and Ce$_{2a}$) can be identified from the agreement between the theoretical
calculation and experimental results. It remains to be seen whether the same method can be applied successfully to other 
phosphors. For this purpose, other materials for which experimental data are known should be analyzed. This should be followed by large-scale computations for new dopant-hosts combinations.

\begin{acknowledgments}
We acknowledge discussions with D. Waroquiers, and thank J.-M. Beuken for computational help. 
This work, done in the framework of ETSF (project number 551),
has been supported by the Fonds de la Recherche Scientifique (FRS-FNRS Belgium) through a
FRIA fellowship (S.P.) and the PdR Grant No. T.0238.13 - AIXPHO.
Computational resources have been provided by the supercomputing facilities
of the Universit\'e catholique de Louvain (CISM/UCL)
and the Consortium des Equipements de Calcul Intensif en
F\'ed\'eration Wallonie Bruxelles (CECI) funded by the FRS-FNRS under Grant No. 2.5020.11.
\end{acknowledgments}

\appendix*
\renewcommand\thefigure{A\arabic{figure}}
\setcounter{figure}{0}
\renewcommand\thetable{A\arabic{table}}
\setcounter{table}{0}   
\section{The Dorenbos model} 
To explain the different optical performance between LaSi$_{3}$N$_{5}$:Ce and
La$_{3}$Si$_{6}$N$_{11}$:Ce$_{2a}$, a qualitative analysis has been performed based on the semi-empirical model proposed by Dorenbos,\cite{Dorenbos2000a,Dorenbos2000b,Dorenbos2002} and the corresponding structural geometry obtained from this work, including the consideration of ground state as well as the excited state.  

\begin{figure}
\includegraphics[scale=0.38]{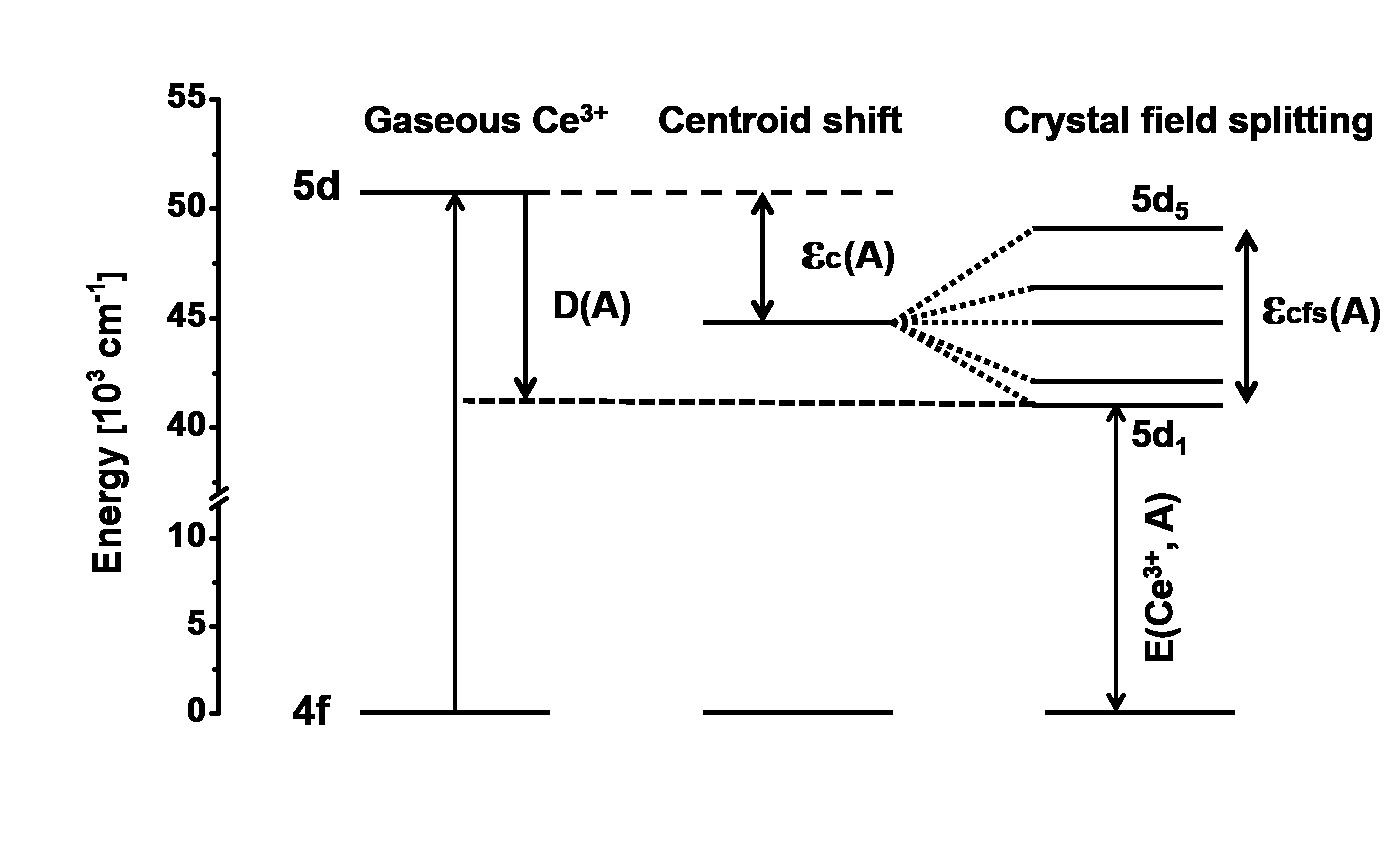}
\caption{Dorenbos model on the energy of [Xe]5d electron configuration of Ce$^{3+}$ ion. The $\varepsilon_{c}$(A), $\varepsilon_{cfs}$(A) and D(A) indicate the centroid shift, crystal field splitting and red-shift of Ce$_{5d}$ energy level in compound A, respectively.\cite{Dorenbos2000b}}\label{figure_A1}
\end{figure} 

First, it is worthy to briefly introduce the Dorenbos model here. Figure \ref{figure_A1} shows the basic idea. The red-shift of the first 4f$\rightarrow$5d transition D(A) in the compound A, can be written as:

\begin{eqnarray}
D(A) &=& \varepsilon_{c}(A) + \dfrac{1}{r(A)}\varepsilon_{cfs}(A) - 1890cm^{-1}
\end{eqnarray}

in which the $\varepsilon_{c}$(A) is the centroid shift of the 5d eneregy relative to the free ion, defined as follows: 

\begin{eqnarray}
\varepsilon_{c}(A) &=& 1.44\times{10}^{17}\Sigma_{i=1}^N\dfrac{\alpha_{sp}^i}{R_{i}^6}
\end{eqnarray}  

In the above formula, $\alpha_{sp}^i$ is the spectroscopic polarization of anion $i$ located at distance of R$_i$ from  the Ce$^{3+}$ ion in the relaxed structure. The summation is over all anions N in the coordinated environment. For nitrides, the qualitative relationship between $\alpha_{sp}^i$ and electronegativity of the cations is demonstrated as:\cite{Wang2015,Wang2016}

\begin{eqnarray}
\alpha_{sp}^N &=& 0.87 + \dfrac{18.76}{\chi_{av}^2}
\end{eqnarray}

where the electronegativity is 

\begin{eqnarray}
\chi_{av} &=& \dfrac{1}{N}\Sigma_{i=1}^M\dfrac{Z_{i}\chi_{i}}{\gamma}
\end{eqnarray}   

This formula is obtained from the reason that a cation of formal charge +Z$_{i}$ will bind on average with Z$_{i}$/$\gamma$ anions of formal charge of -$\gamma$. The summation is over all cations M in the compound, and N is the number of anions.\citep{Dorenbos2002,Wang2015} 

Another parameter affecting the spectroscopic red-shift is the contribution from the crystal field shift, $\dfrac{1}{r(A)}\varepsilon_{cfs}$(A). The crystal-field splitting $\varepsilon_{cfs}$(A) is defined as the energy difference between the lowest and highest 5d level. A fraction 1/r(A) contributes to the red-shift, where r(A) usually varies between 1.7 and 2.4. The $\varepsilon_{cfs}$(A) is determined as 

 \begin{eqnarray}
\varepsilon_{cfs} &=& \dfrac{\beta}{R_{av}^2}
\end{eqnarray}     

in which $\beta$ is a parameter related to the shape and size of the anion polyhedron coordinated to the Ce$^{3+}$ ion, and R$_{av}$ is average distance between the Ce$^{3+}$ ion and anions in the relaxed structure.

In our work, the Dorenbos model has been used to analyse the different luminescent behaviour of LaSi$_{3}$N$_{5}$:Ce and
La$_{3}$Si$_{6}$N$_{11}$:Ce.

\newpage
\bibliography{LSN} 

\end{document}